# Deadlock-Free Typestate-Oriented Programming

Luca Padovani[a]

a Dipartimento di Informatica, Università di Torino, Italy

**Abstract** *Context.* TypeState-Oriented Programming (TSOP) is a paradigm intended to help developers in the implementation and use of mutable objects whose public interface depends on their private state. Under this paradigm, well-typed programs are guaranteed to conform with the protocol of the objects they use.
*Inquiry.* Previous works have investigated TSOP for both sequential and concurrent objects. However, an important difference between the two settings still remains. In a sequential setting, a well-typed program either progresses indefinitely or terminates eventually. In a concurrent setting, protocol conformance is no longer enough to avoid *deadlocks*, a situation in which the execution of the program halts because two or more objects are involved in mutual dependencies that prevent any further progress.
*Approach.* In this work, we put forward a refinement of TSOP for concurrent objects guaranteeing that well-typed programs not only conform with the protocol of the objects they use, but are also deadlock free. The key ingredients of the type system are *behavioral types*, used to specify and enforce object protocols, and *dependency relations*, used to represent abstract descriptions of the dependencies between objects and detect circularities that might cause deadlocks.
*Knowledge.* The proposed approach stands out for two features. First, the approach is fully compositional and therefore scalable: the objects of a large program can be type checked in isolation; deadlock freedom of an object composition solely depends on the types of the objects being composed; any modification/refactoring of an object that does not affect its public interface does not affect other objects either. Second, we provide the first deadlock analysis technique for *join patterns*, a high-level concurrency abstraction with which programmers can express complex synchronizations in a succinct and declarative form.
*Grounding.* We detail the proposed typing discipline for a core programming language blending concurrent objects, asynchronous message passing and join patterns. We prove that the type system is sound and give non-trivial examples of programs that can be successfully analyzed. A Haskell implementation of the type system that demonstrates the feasibility of the approach is publicly available.
*Importance.* The static analysis technique described in this work can be used to certify programs written in a core language for concurrent TSOP with proven correctness guarantees. This is an essential first step towards the integration and application of the technique in a real-world developer toolchain, making programming of such systems more productive and less frustrating.

## 1 Intoduction

The public interface of a mutable object may depend on its private state [6]. A typical example is that of a file, which can be read or written if it is open, but not if it is closed. Yet, in most programming languages the interface of files does not specify any formal correlation between the availablity of read/write methods and the state of the file on which they are invoked. At best, the textual description of these methods warns the programmer about the intended *usage protocol* of files. Typestate-oriented programming [17, 2, 42, 21] (TSOP for short) provides linguistic constructs and types that facilitate the implementation and use of objects with structured protocols, enabling the detection of protocol violations at compile time.

TSOP uses *control-flow analysis* to track the state of objects and to determine whether a certain method invocation in a specific point of the program is valid or not. This technique falls short for *concurrent objects*, those objects that may be accessed and modified unpredictably by several concurrent processes. To overcome this difficulty, Crafa and Padovani [16] have proposed a hybrid approach to concurrent TSOP that combines compile-time checks and execution-time synchronizations. A significant gap between the sequential and concurrent settings still remains, though. In a sequential setting, the fact that a program conforms with the protocol of the objects it uses is enough to guarantee that the program either progresses indefinitely or terminates eventually. In a concurrent setting, this is no longer the case: there are programs that conform with the protocol of the objects they use and yet their execution stops prematurely because a deadlock has occurred. The aim of this work is to refine the concurrent approach to TSOP put forward by Crafa and Padovani so as to prevent these situations.

We see an instance of the problem in the following program, which creates a `future` variable [4] and its `user` in an object-based language that supports concurrent TSOP and uses asynchronous message passing as the sole interaction mechanism [16]:

```
1 new future : #FutureT
2 [ EMPTY       & Resolve(n) ▶ future!RESOLVED(n)
3 | RESOLVED(n) & Get(r)     ▶ future!RESOLVED(n) & r!Reply(n) ]
4 in future!EMPTY
5 &
6 new user : #UserT
7 [ READ(future)             ▶ user!WRITE(future) & future!Get(user)
8 | WRITE(future) & Reply(n) ▶ user!DONE          & future!Resolve(n) ]
9 in user!READ(future)
```

The future variable (lines 1–4) can be in one of two states. When it is `EMPTY`, `future` accepts a `Resolve` message carrying some value `n` and becomes `RESOLVED`, storing `n` in its state (line 2). When it is `RESOLVED`, `future` accepts a `Get` message carrying a reference `r` to another object, sends the stored value `n` to `r` and remains in the `RESOLVED` state (line 3). The future variable is `EMPTY` when initialized (line 4). The `user` object (lines 6–9) can be in one of three states. When in state `READ`, `user` sends a request to retrieve the content of `future` (a reference to a future variable stored in `user`'s state) and spontaneously moves into the `WRITE` state (line 7). When in state

`WRITE`, `user` accepts a `Reply` message carrying a value `n`, resolves `future` with `n` and moves into the `DONE` state (line 8). The object `user` is initialized in the `READ` state (line 9) and, once in state `DONE`, becomes inert (it has no transitions from this state).

Since it is impossible for `user` to retrieve a value from `future` *before* `future` has been resolved, this program will deadlock. Crafa and Padovani [16] have shown that *commutative regular expressions* [12] can serve as natural type specifications for concurrent object protocols: a type describes the set of legal message configurations that can be targeted to objects with that type; commutativity of the connective · captures the fact that the order of concurrent message outputs is unpredicatble and therefore irrelevant. As an example, a sensible type `#FutureT` for `future` is

$$(\texttt{EMPTY} \cdot \texttt{Resolve} + \texttt{RESOLVED}) \cdot {*}\texttt{Get}$$

which specifies that `future` is always either `EMPTY` or `RESOLVED`, that it *must* receive exactly one `Resolve` message when it is `EMPTY`, and that it *may* receive any number of `Get` messages regardless of its state. According to the type system of Crafa and Padovani, `user` conforms with `#FutureT` and the above program is well typed. Unfortunately, the possibility of sending `Get` messages to `future` not knowing `future`'s state is exploited to generate a *circular dependency* between `future` and `user` that leads to a deadlock. To prevent deadlocks like this, we put forward an approach that combines two complementary mechanisms. We keep using *behavioral types* such as `#FutureT` to specify concurrent object protocols and impose restrictions and obligations on their users. In addition, we track the *dependencies* between the objects to detect potentially dangerous circularities. In the above program, the *two* outputs `user!WRITE(future)` and `future!Get(user)` on line 7 hint at such a circularity between `user` and `future`.

Deadlocks in concurrent programming have been a longstanding issue and many approaches aimed at deadlock prevention have been proposed. We will give a detailed account of them in Section 6 but we anticipate that, among these works, our proposal stands out for two features. First of all, it is *compositional*: deadlock freedom of a compound system can be established solely using information gathered from its sub-systems in isolation. It is never the case that, to successfully combine two well-typed sub-systems, the sub-systems must be re-analyzed or re-typed differently. The analysis is necessarily conservative (deadlock freedom is undecidable in general), but expressive enough to handle non-trivial programs. Second, our approach is the first one that attacks the problem of deadlock analysis for *concurrent objects with join patterns* [18, 19, 8, 40, 41, 22, 43]. Actors [24, 1, 23] are a special case of such objects.

**Structure of the paper** We recall the formal model for concurrent TSOP [19, 16] and formalize the notions of *protocol-conformant* and *deadlock-free* processes (Section 2). Then, we walk through a series of simple examples to illustrate the key ideas of our analysis technique (Section 3) before presenting the typing rules and their soundness properties (Section 4). In the latter parts of the paper we discuss a few more complex examples (Section 5) and related work (Section 6). Proofs and additional technical material are in Appendix A. All the code in shaded background can be type checked and possibly executed using Cobalt**Blue** [34], a publicly available implementation of the presented type system.

**Table 1** Syntax of the behaviorally typed Objective Join Calculus.

| | | | |
|---|---|---|---|
| **Process** | $P,Q$ | ::= | $\texttt{done} \mid u\,!\,M \mid P \mathbin{\&} Q \mid \texttt{new}\ a : t = [C]\ \texttt{in}\ P$ |
| **Molecule** | $M,N$ | ::= | $\mathtt{m}(\overline{u}) \mid M \mathbin{\&} N$ |
| **Pattern** | $J,K$ | ::= | $\mathtt{m}(\overline{x}) \mid J \mathbin{\&} K$ |
| **Class** | $C,D$ | ::= | $J \blacktriangleright P \mid C \mid D$ |
| **Type** | $t,s$ | ::= | $\mathbb{0} \mid \mathbb{1} \mid \mathtt{m}(\overline{t}) \mid t+s \mid t \cdot s \mid *t$ |

## 2 A Formal Model for Concurrent Typestate-Oriented Programming

Following Crafa and Padovani [16] we use a typed version of the Objective Join Calculus [19] as formal model for concurrent TSOP. We assume an infinite set of *object names* ranged over by $a$, $b$, and a disjoint, infinite set of *variables* ranged over by $x$, $y$. We let $u$, $v$ range over *names*, which are either object names or variables. We also assume a set of *message tags* ranged over by $\mathtt{m}$. We write $\overline{e}$ or $e_1,\dots,e_n$ for *sequences* of various entities and say that $n$ is the length of the sequence. For example, $\overline{u}$ stands for the sequence of names $u_1,\dots,u_n$. We write $\langle e_1,\dots,e_n\rangle$ for *multisets* of entities. When no ambiguity may arise, we occasionally drop the parentheses $\langle \cdots \rangle$ and let the sequence $e_1,\dots,e_n$ denote the corresponding multiset $\langle e_1,\dots,e_n\rangle$.

The syntax of the Objective Join Calculus is shown in Table 1 and comprises processes, molecules, patterns, classes and types. The term $\texttt{done}$ denotes the terminated process, which does nothing. The term $u\,!\,M$ denotes the process that sends the molecule $M$ to the object $u$. A *molecule* is a composition $\mathtt{m}_1(\overline{u}_1) \mathbin{\&} \cdots \mathbin{\&} \mathtt{m}_k(\overline{u}_k)$ of messages. Each message $\mathtt{m}(\overline{u})$ consists of a tag $\mathtt{m}$ and a sequence of arguments $u_1,\dots,u_n$. We say that $n$ is the *arity* of the message and we just write $\mathtt{m}$ instead of $\mathtt{m}(\,)$ when $n = 0$. Note that a process $u\,!\,M$ has no continuation, meaning that the calculus is asynchronous. Sequential composition is modeled by means of explicit continuation passing. The term $P \mathbin{\&} Q$ denotes the parallel composition of $P$ and $Q$. The term $\texttt{new}\ a : t = [C]\ \texttt{in}\ P$ denotes a process that creates an object $a$ of type $t$ and class $C$ and whose scope is $C$ and $P$. The type of an object specifies how the object is supposed to be used. Syntax and semantics of types will be given later. A *class* consists of a finite set of *reaction rules* $J_1 \blacktriangleright P_1 \mid \cdots \mid J_n \blacktriangleright P_n$ and determines how the object reacts to the messages targeted to it. Each reaction $J \blacktriangleright P$ is made of a pattern $J$ and a process $P$. When a molecule of messages that matches the pattern $J$ is targeted to the object, the molecule is atomically consumed and the corresponding process $P$ is spawned.

The definitions of free and bound names for processes are standard and omitted for space reasons [19]. Hereafter we will consider processes equal modulo bound names.

The semantics of the Objective Join Calculus is akin to a chemical process [9] whereby the state of the computation is a *solution* $\mathscr{D} \Vdash \mathscr{P}$ consisting of a finite map $\mathscr{D}$ from object names to object definitions and a multiset (or "soup") $\mathscr{P}$ of molecules and processes. The map $\mathscr{D}$ determines which chemical reactions may occur in $\mathscr{P}$. Chemical reactions consume molecules matching certain patterns from $\mathscr{P}$ and produce new processes in $\mathscr{P}$. The molecules and processes in $\mathscr{P}$ disgregate and recombine depending on the temperature of the solution. Reflexivity [18] allows new chemical reactions to be added to $\mathscr{D}$ dynamically, as the computation progresses.

**Table 2** Semantics of the Objective Join Calculus.

| | | | | |
|---|---|---|---|---|
| [DONE] | $\Vdash \texttt{done}$ | $\rightleftharpoons$ | $\Vdash$ | |
| [NEW] | $\mathscr{D} \Vdash \texttt{new}\ a : t = [C]\ \texttt{in}\ P, \mathscr{P}$ | $\rightleftharpoons$ | $\mathscr{D}, a : t = C \Vdash P, \mathscr{P}$ | $a \notin \mathsf{fn}(\mathscr{P})$ |
| [PAR] | $\Vdash P \mathbin{\&} Q$ | $\rightleftharpoons$ | $\Vdash P, Q$ | |
| [JOIN] | $\Vdash a\,!(M \mathbin{\&} N)$ | $\rightleftharpoons$ | $\Vdash a\,!M, a\,!N$ | |
| [RED] | $a : t = C \Vdash a\,!\sigma J$ | $\rightarrow$ | $a : t = C \Vdash \sigma P$ | $J \blacktriangleright P \in C$ |

More precisely, the semantics of the Objective Join Calculus is given by the relations $\rightharpoonup$, $\rightharpoondown$ and $\rightarrow$ defined in Table 2. As customary for the Objective Join Calculus [19], in the table we only show the components of the solution that are affected by each relation to avoid clutter. For example, rule [DONE] as shown in Table 2 actually stands for $\mathscr{D} \Vdash \texttt{done}, \mathscr{P} \rightleftharpoons \mathscr{D} \Vdash \mathscr{P}$. The relations $\rightharpoonup$ and $\rightharpoondown$ change the temperature of the solution by heating and cooling it, respectively. These trasformations are reversible and defined by the four topmost rules. The rule [DONE] states that `done` terms evaporate and condensate, as the temperature changes. The rule [NEW] moves object definitions between processes and definitions. The side condition $a \notin \mathsf{fn}(\mathscr{P})$ makes sure that no occurrence of $a$ is accidentally captured (when the solution heats up) and, conversely, that no occurrence of $a$ becomes free (when the solution cools down). Given that we have assumed an infinite supply of object names and that we silently rename bound names in processes, it is always possible to perform both transformations. The rules [PAR] and [JOIN] break and recompose processes and molecules. In the latter case, molecules must be targeted to the same object $a$. There is only one reduction rule [RED] representing a non-reversible chemical reaction, which occurs when the solution contains a molecule $a\,!M$ targeted to some object $a$ and $M$ matches the left-hand side $J$ of a reaction $J \blacktriangleright P$ in the class of $a$. The matching is witnessed by a substitution $\sigma$ from variables to names such that $\sigma J = M$. In this case, the reaction *fires*: the molecule $M$ is atomically removed from the solution and replaced by the process $P$ on the right-hand side of the reaction with its free variables substituted according to $\sigma$.

**Example 1.** We may use these definitions to run the program discussed in Section 1. Initially, the only available messages are `future!EMPTY` (line 4) and `user!READ(future)` (line 9). The second message triggers the first reaction of `user` (line 7) which generates two further messages `user!WRITE(future)` and `future!Get(user)`. At this point, no combination of the available messages is capable of triggering any reaction, therefore the execution halts. ■

Types specify *how* objects can be *used* and *shared*. Using an object means targeting the object with one or more messages. There may be messages that are mutually exclusive, so that only one of them can be chosen and sent, and others that can be sent concurrently. There may be objects that can only be used by one process at a time and others that can be shared. There are objects that *must* be used and others that can be discarded. We specify these possibilities, obligations and prohibitions using a language of behavioral types akin to regular expressions, but where composition $\cdot$ is *commutative* (Table 1).

The type $\mathbb{0}$ describes absurd objects: there is no legal way of using an object with this type, and yet not using the object is disallowed as well. While this type is literally useless for the programmer, its role is important in the theory of types that we will develop and cannot be omitted altogether from the syntax. As we will see, $\mathbb{0}$ is the top element of the type hierarchy and may be generated by a crucial operator on types. The type $\mathbb{1}$ describes objects that can only be discarded. Any other usage (like sending a message to the object) is forbidden. The *message type* $\mathtt{m}(\overline{t})$ describes an object that must be used as target for a single message with tag $\mathtt{m}$ and arguments of type $\overline{t}$. A process owning a reference to an object with this type has the obligation to use the object as prescribed. Simply discarding the object is disallowed. We write $\mathtt{M}$ for an arbitrary message type $\mathtt{m}(\overline{t})$, we say that the length of $\overline{t}$ is the *arity* of $\mathtt{M}$ and we write just $\mathtt{m}$ instead of $\mathtt{m}()$ when $\overline{t}$ is the empty sequence. Next we have three connectives for building compound types. The sum $+$ represents *choice*: an object of type $t+s$ must be used *either* as specified by $t$ *or* as specified by $s$. For example, a process owning a reference to an object of type $\mathtt{a}+\mathtt{b}$ must send either an $\mathtt{a}$ message or a $\mathtt{b}$ message to the object. Not using the object or sending both $\mathtt{a}$ and $\mathtt{b}$ messages are disallowed. On the other hand, an object of type $\mathbb{1}+\mathtt{m}$ may be either discarded (if "used" as specified by $\mathbb{1}$) or used as target for an $\mathtt{m}$ message. The product $\cdot$ represents *concurrency*: an object of type $t \cdot s$ must be used *both* as specified by $t$ *and also* as specified by $s$, by possibly concurrent processes owning references to it. For example, a process owning a reference to an object of type $\mathtt{a}\cdot\mathtt{b}$ must send both an $\mathtt{a}$ message and a $\mathtt{b}$ message to the object. Sending only either or none of them is disallowed. The exponential $*$ represents *unrestricted, possibly concurrent* usage: an object of type $*t$ may be used any number of times by possibly concurrent processes, each time as specified by $t$. For example, a process owning a reference to an object of type $*(\mathtt{a}\cdot\mathtt{b})$ may send any number of $\mathtt{a}$ and $\mathtt{b}$ messages to the object, provided that the number of 'a' messages is the same as the number of 'b' messages. In addition to these types, in the examples we occasionally use `int` or `#Number` to describe basic values.

To specify possibly infinite protocols, we interpret the productions for types in Table 1 coinductively. In other words, we consider as types the possibly infinite trees generated by those productions. We impose two restrictions on such type trees, namely *regularity* and *contractiveness*. Regularity requires that each type tree is made of finitely many distinct subtrees. This guarantees that the type is finitely representable, either with the well-known $\mu$-notation for recursive types or by means of a finite system of equations [15]. Contractiveness requires that each infinite branch of a tree type must go through infinitely many message types, so as to avoid types such as those satisfying the equation $t = t + t$ or $t = *t$, which do not provide any information.

We formalize the semantics of a type by means of its valid configurations, describing which combinations of messages can be sent concurrently to an object with that type:

**Definition 1** (valid configurations). *Configurations*, ranged over by A, B, are multisets of message types. The *valid configurations* of a type $t$, denoted by $[\![t]\!]$, are defined by

$$[\![\mathbb{0}]\!] \stackrel{\text{def}}{=} \emptyset \qquad [\![t+s]\!] \stackrel{\text{def}}{=} [\![t]\!] \cup [\![s]\!] \qquad [\![\mathtt{M}]\!] \stackrel{\text{def}}{=} \{\langle \mathtt{M} \rangle\}$$
$$[\![\mathbb{1}]\!] \stackrel{\text{def}}{=} \{\langle\rangle\} \qquad [\![t \cdot s]\!] \stackrel{\text{def}}{=} \{\mathsf{A} \uplus \mathsf{B} \mid \mathsf{A} \in [\![t]\!] \wedge \mathsf{B} \in [\![s]\!]\} \qquad [\![{*t}]\!] \stackrel{\text{def}}{=} \bigcup_{i\in\mathbb{N}} [\![t^i]\!]$$

where $t^0 \stackrel{\text{def}}{=} \mathbb{1}$ and $t^{i+1} \stackrel{\text{def}}{=} t \cdot t^i$ and $\uplus$ denotes multiset union.

Contractiveness makes sure that the set of valid configurations is well defined, since there is never any need to peek inside message types. As an exampe, we have $[\![\mathtt{a} \cdot (\mathtt{b} + \mathtt{c})]\!] = \{\langle \mathtt{a}, \mathtt{b}\rangle, \langle \mathtt{a}, \mathtt{c}\rangle\}$ meaning that an object of type $\mathtt{a} \cdot (\mathtt{b} + \mathtt{c})$ must be used as target for either an 'a' and a 'b' message or an 'a' and a 'c' message. It follows that the types $\mathtt{a} \cdot (\mathtt{b} + \mathtt{c})$ and $\mathbb{1} \cdot \mathtt{a} \cdot (\mathtt{b} + \mathtt{c})$ and $\mathtt{a} \cdot \mathtt{b} + \mathtt{a} \cdot \mathtt{c}$ have the same valid configurations and indeed they will be equivalent according to Definition 2. As another example, we have $[\![{*}(\mathtt{a} \cdot \mathtt{b})]\!] = [\![\mathbb{1}]\!] \cup [\![\mathtt{a} \cdot \mathtt{b}]\!] \cup [\![\mathtt{a} \cdot \mathtt{a} \cdot \mathtt{b} \cdot \mathtt{b}]\!] \cup \cdots = \{\langle\rangle, \langle \mathtt{a}, \mathtt{b}\rangle, \langle \mathtt{a}, \mathtt{a}, \mathtt{b}, \mathtt{b}\rangle, \dots\}$. Notice the difference between $\mathbb{0}$ and $\mathbb{1}$. The only valid configuration of the latter is $\langle\rangle$, meaning that not sending any message to an object of type $\mathbb{1}$ is allowed. On the other hand, $\mathbb{0}$ has no valid configurations at all, so there is no way to conform with it.

We are now ready to introduce subtyping:

**Definition 2** (subtyping). Let $\leqslant$ be the largest relation on types such that $t \leqslant s$ and $\langle \mathtt{m}_i(\overline{s_i})\rangle_{i\in I} \in [\![s]\!]$ imply $\langle \mathtt{m}_i(\overline{t_i})\rangle_{i\in I} \in [\![t]\!]$ and $\overline{s_i \leqslant t_i}$ for every $i \in I$. We say that $t$ is a *subtype* of $s$ (and $s$ a *supertype* of $t$) if $t \leqslant s$ holds. We say that $t$ and $s$ are equivalent, written $t \simeq s$, iff $t \leqslant s$ and $s \leqslant t$.

Subtyping is best understood bearing in mind the usual safe substitution principle: if $t \leqslant s$, then it is safe to replace an object of type $s$ with an object of type $t$. For example, we have $\mathtt{a} + \mathtt{b} \leqslant \mathtt{a}$: a process using an object of type $\mathtt{a}$ can (and must) send an 'a' message to the object. This is also a valid usage of an object of type $\mathtt{a} + \mathtt{b}$, which allows sending either an 'a' message or a 'b' message to the object. As usual, subtyping is contravariant on argument types. For example, we have $\mathtt{m}(\mathtt{a}) \leqslant \mathtt{m}(\mathtt{a}+\mathtt{b})$. A process using an object of type $\mathtt{m}(\mathtt{a} + \mathtt{b})$ will send a message $u(v)$ to $u$ and $u$ will use $v$ for sending either an $\mathtt{a}$ message or a $\mathtt{b}$ message. Then, it is safe to send the same message also to an object of type $\mathtt{m}(\mathtt{a})$, which will deterministically send an $\mathtt{a}$ message to $v$. The properties of $\leqslant$ that have been exemplified so far are standard for subtyping relations on object types. The original ingredient of $\leqslant$ is that it also accounts for message combinations. For example, $\mathtt{a} \cdot (\mathtt{b} + \mathtt{c}) \leqslant \mathtt{a} \cdot \mathtt{c}$, since using an object according to the type $\mathtt{a} \cdot \mathtt{c}$ is one of the allowed usages of an object of type $\mathtt{a} \cdot (\mathtt{b} + \mathtt{c})$. In general, it is possible to show that $\leqslant$ includes all the laws of Commutative Kleene Algebra [12]: both $+$ and $\cdot$ are commutative and respectively have $\mathbb{0}$ and $\mathbb{1}$ as neutral elements, $+$ is idempotent, $\cdot$ distributes over $+$ and is absorbed by $\mathbb{0}$. Other notable laws include ${*t} \leqslant {*t} \cdot {*t}$ and ${*t} \leqslant t$, which allow an object of type ${*t}$ to be shared and used according to $t$ by an arbitrary number of processes. Finally, $\leqslant$ is a pre-congruence with respect to all type connectives [16].

We give characterizations of "usable" types – those describing objects that *can* be used – and "relevant" types – those describing objects that *must* be used.

**Definition 3.** We say that $t$ is *usable* if $\mathbb{0} \not\leqslant t$ and that it is *relevant* if $t \not\leqslant \mathbb{1}$.

A usable type has at least one (possibly empty) valid configuration and a relevant type does not have $\langle\rangle$ among its valid configurations. The type $\mathbb{0}$ is relevant but unusable and $\mathbb{1}$ is usable but irrelevant. Types like $\mathtt{m}$ or $\mathtt{a}+\mathtt{b}$ or $\mathtt{a}\cdot\mathtt{b}$ are both usable and relevant. We extend this terminology from types to the objects they describe.

The next notion we introduce is a "derivative" operator allowing us to compute the residual of a type when we remove one or more message types with a given tag. This operator is closely related to, and takes its name from, Brzozowski's derivative [10].

**Definition 4** (type derivative)**.** We write $\mathtt{M} \approx \mathtt{M}'$ if $\mathtt{M}$ and $\mathtt{M}'$ have the same tag. The *derivative* of a type $t$ with respect to a message type $\mathtt{M}$, written $t[\mathtt{M}]$, is defined by:

$$\mathbb{0}[\mathtt{M}] = \mathbb{1}[\mathtt{M}] \stackrel{\text{def}}{=} \mathbb{0} \qquad \mathtt{M}'[\mathtt{M}] \stackrel{\text{def}}{=} \begin{cases} \mathbb{1} & \text{if } \mathtt{M} \approx \mathtt{M}' \\ \mathbb{0} & \text{otherwise} \end{cases} \qquad (t+s)[\mathtt{M}] \stackrel{\text{def}}{=} t[\mathtt{M}] + s[\mathtt{M}]$$
$$(*t)[\mathtt{M}] \stackrel{\text{def}}{=} t[\mathtt{M}] \cdot *t \qquad (t\cdot s)[\mathtt{M}] \stackrel{\text{def}}{=} t\cdot s[\mathtt{M}] + t[\mathtt{M}]\cdot s$$

The derivative of $t$ with respect to $\mathsf{A} = \langle \mathtt{M}_1, \dots, \mathtt{M}_n\rangle$ is $t[\mathsf{A}] \stackrel{\text{def}}{=} t[\mathtt{M}_1]\cdots[\mathtt{M}_n]$.[1]

If $t$ is the protocol of an object, then $t[\mathtt{m}]$ is the residual protocol of the same object after an $\mathtt{m}$ message has been sent to it. For example, if $t = \mathtt{a}\cdot(\mathtt{b}+\mathtt{c})$ and we send an 'a' message to the object, in order to conform with $t$ we are still supposed to use the object according to the type $t[\mathtt{a}] = \mathtt{a}\cdot(\mathtt{b}+\mathtt{c})[\mathtt{a}] + \mathtt{a}[\mathtt{a}]\cdot(\mathtt{b}+\mathtt{c}) \simeq \mathtt{b}+\mathtt{c}$. If instead we send a 'd' message to the same object of type $t$, then we obtain $t[\mathtt{d}] \simeq \mathbb{0}$. By sending 'd' we have violated the protocol $t$ and now we are left with an unusable object. In general, usability and relevance help us formalize conformance with, completion and violation of an object's protocol. If we have sent to an object of type $t$ a multiset of messages described by the configuration $\mathsf{A}$ and $t[\mathsf{A}]$ is still usable, then we conformed with the object's protocol. If $t[\mathsf{A}]$ is also irrelevant, then no more duties remain and we have completed the object's protocol. On the contrary, if $t[\mathsf{A}]$ is unusable, then we have not used the object appropriately. Concerning protocol conformance, we have:

**Definition 5** (conformant process)**.** We say that $P$ is a *conformant process* if, whenever $\emptyset \Vdash P \Longrightarrow \mathscr{D}, a : t = C \Vdash \langle a\,!\,\mathtt{m}_i(\overline{c_i})\rangle_{i\in I}, \mathscr{P}$, then $t[\langle \mathtt{m}_i\rangle_{i\in I}]$ is usable and, if $t[\langle \mathtt{m}_i\rangle_{i\in I}]$ is also relevant, then $a$ occurs in $\mathscr{P}$ and/or in one of the $\overline{c_i}$ for some $i \in I$.

In words, if a conformant process $P$ sends some configuration $\langle \mathtt{m}_1, \dots, \mathtt{m}_n\rangle$ of messages to an object $a$ with type $t$, then $t[\langle \mathtt{m}_1, \dots, \mathtt{m}_n\rangle]$ is usable, meaning that there is a valid configuration of $t$ that includes (but is not necessarily equal to) $\langle \mathtt{m}_1, \dots, \mathtt{m}_n\rangle$. That is, $P$ has not violated the protocol of $a$. Moreover, if $t[\langle \mathtt{m}_1, \dots, \mathtt{m}_n\rangle]$ is also relevant, meaning that there are pending obligations with respect to $a$, then there is at least one reference to $a$ in the system that may be used for this purpose.

Concerning protocol completion, and therefore deadlock freedom, we have:

**Definition 6** (deadlock-free process)**.** We say that $P$ is a *deadlock-free process* if, whenever $\emptyset \Vdash P \Longrightarrow \mathscr{D}, a : t = C \Vdash \langle a\,!\,\mathtt{m}_i(\overline{c_i})\rangle_{i\in I}, \mathscr{P}\ (\rightharpoonup \cup \rightarrow)^* \not\rightarrow$ and there are no messages targeted to $a$ in $\mathscr{P}$, then $t[\langle \mathtt{m}_i\rangle_{i\in I}]$ is irrelevant.

[1] Strictly speaking, $t[\langle \mathtt{M}_1, \dots, \mathtt{M}_n\rangle]$ depends on the order in which the $\mathtt{M}_i$'s are considered, but it can be shown that all the possible outcomes are equivalent, *cf.* Lemma 2.

In words, if $\emptyset \Vdash P$ reduces to a state from which no further reductions are possible and we consider the overall configuration $\langle \mathtt{m}_1, \ldots, \mathtt{m}_n \rangle$ of messages targeted to an object $a$ of type $t$, then $t[\langle \mathtt{m}_1, \ldots, \mathtt{m}_n \rangle]$ is irrelevant, meaning that $\langle \mathtt{m}_1, \ldots, \mathtt{m}_n \rangle$ is a valid configuration of $t$. That is, $P$ has no unfulfilled obligations with respect to $a$.

**Example 2.** Consider again the program in Section 1 and recall from Example 1 that its execution halts in a state consisting of the messages `future!EMPTY`, `future!Get(user)` and `user!WRITE(future)`. We have $\mathtt{\#FutureT}[\langle \mathtt{EMPTY}, \mathtt{Get} \rangle] \simeq \mathtt{Resolve} \cdot {*}\mathtt{Get}$ which is usable and relevant. Therefore, the program is conformant but not deadlock free. ■

Note that this definition of deadlock freedom is more general than the mere absence of junk messages. Specifically, Definiton 6 tolerates the presence unconsumed messages targeted to $a$ if this is allowed by its type $t$. This notion of deadlock freedom accounts for the presence of state messages, such as `RESOLVED` in the case of `future`, which need not be consumed in case no more users of the object remain (see Example 3). On the other hand, we can obtain a particular instance of Definition 6 which is easier to understand by taking $I = \emptyset$. In this case, deadlock freedom translates as the property that, if the system halts and no messages are targeted to an object $a$ of type $t$, then it is because the empty configuration of messages is valid for $t$.

## 3 Examples

In this section we use a series of simple programs to introduce the notion of *object dependency*, the key mechanism at the base of our analysis technique. To keep the discussion focused, we take the stance that the analysis should flag those programs that leave unconsumed messages (which we call "junk") when their execution halts. As we have remarked at the end of the previous section, this is only a particular case of undesirable program behavior captured by Definition 6.

The first program we consider concerns an object `obj` with a single reaction that fires when `obj` receives both an `A` message and a `B` message:

```
new obj : *(A·B) [ A & B ▸ done ] in obj!A
```

This program is unable to make any progress because there is no `B` message that can form a molecule with the `A` message and trigger the only reaction of `obj`. As a consequence, the `A` message cannot be consumed and becomes junk. Conversely, in

```
new obj : *(A·B) [ A & B ▸ done ] in obj!A & obj!B & obj!B
```

there are too many `B` messages. One of them can combine with `A` and be consumed by the reaction, but the other one becomes junk. Neither of these two programs is conformant according to Definition 5 because their final state contains a `B` message, the type $({*}(\mathtt{A} \cdot \mathtt{B}))[\mathtt{B}] \simeq \mathtt{A} \cdot {*}(\mathtt{A} \cdot \mathtt{B})$ is relevant, and no other occurrence of `obj` exists in the program. Indeed, they are both ill typed according to the typing discipline of Crafa and Padovani [16], which ensures conformance of well-typed programs.

Unfortunately, conformance alone does not always suffice to prevent junk messages. An example of conformant but deadlocked program is shown below:

```
new obj₁ : *(A·B(A)) [ A & B(x) ▸ x!A ] in
new obj₂ : *(A·B(A)) [ A & B(y) ▸ y!A ] in obj₁!B(obj₂) & obj₂!B(obj₁)
```

Here we have two objects $\mathtt{obj}_1$ and $\mathtt{obj}_2$, each reacting to a molecule consisting of one `A` message and one `B` message. There are two `B` messages respectively targeted to $\mathtt{obj}_1$ and $\mathtt{obj}_2$ and, from a syntactic viewpoint, also two `A` messages targeted to the same objects via the outputs `y!A` and `x!A`, respectively. However, the program is deadlocked according to Definition 6 because neither reaction can fire and $(*(\mathtt{A}\cdot\mathtt{B}))[\mathtt{B}]$ is relevant. The origin of the problem can be traced to the code following the last `in`, where $\mathtt{obj}_2$ is the argument of a message targeted to $\mathtt{obj}_1$, meaning that the obligation to send an `A` message through this occurrence of $\mathtt{obj}_2$ may not be fulfilled until $\mathtt{obj}_1$'s reaction fires. Symmetrically, $\mathtt{obj}_1$ is the argument of a message targeted to $\mathtt{obj}_2$, meaning that the obligation to send an `A` message through this occurrence of $\mathtt{obj}_1$ may not be fulfilled until $\mathtt{obj}_2$'s reaction fires. It is this *mutual dependency* between $\mathtt{obj}_1$ and $\mathtt{obj}_2$ that prevents the program from making any progress. What we learn from this example is that the arguments of a message depend on the target of the very same message and the presence of mutual dependencies may result into a deadlock.

A somewhat degenerate case of mutual dependency is shown in the program

```
new obj : *(A·B(A)) [ A & B(x) ▸ x!A ] in obj!B(obj)
```

where the `A` message that is needed to fire the only reaction of `obj` is produced by the reaction itself. Here we see that a *self-dependency* arises if the same object (`obj` in this case) is both the target and an argument of the same message.

Regrettably, not all self-dependencies are so easy to spot. Sometimes they arise as the result of more complex interactions. For example,

```
new obj₁ : *(A·B(A))   [ A & B(x) ▸ x!A    ] in
new obj₂ : *C(B(A),A) [ C(x,y)   ▸ x!B(y) ] in obj₂!C(obj₁, obj₁)
```

leads to a self-dependency on $\mathtt{obj}_1$ by the firing of the reduction of $\mathtt{obj}_2$, when $\mathtt{obj}_1$ is substituted in place of both `x` and `y`. Yet, `x` and `y` are syntactically different variables in the source code of the program. The suspicious trait in this example is the fact that $\mathtt{obj}_1$ occurs twice as argument in the same message. Preventing the same object to occur more than once in a single message is a common strategy at the heart of several type systems ensuring the absence of deadlocks [11, 44, 31]. Unfortunately, the availability of join patterns that sets the Objective Join Calculus apart from other models of communicating processes renders this strategy ineffective. For example

```
new obj₁ : *(A·B(A))       [ A    & B(x) ▸ x!A    ] in
new obj₂ : *(C(B(A))·D(A)) [ C(x) & D(y) ▸ x!B(y) ] in
obj₂!C(obj₁) & obj₂!D(obj₁)
```

leads to the same self-dependency on $\mathtt{obj}_1$ as before, except that here $\mathtt{obj}_1$ is never used twice in the same message. In this case, the only hint at a potential problem is that the *same* dependency (of $\mathtt{obj}_1$ on $\mathtt{obj}_2$) arises *twice*. In summary, the multiplicity of dependencies – and not just the presence or lack thereof – is relevant, even if these dependencies do not form cycles.

Our type system flags as ill typed all the programs shown in this section by combining the enforcement of *object protocols* with a mechanism that tracks *object dependencies*. It is now time to look at the acutal typing rules and this mechanism in greater detail.

## 4 The Type System

As usual we need type environments for tracking the type of free names when typing processes, patterns and molecules. We introduce the corresponding notation below.

**Definition 7** (type environment). A *type environment* $\Gamma$ is a partial map from names to types equivalently written $u_1 : t_1, \dots, u_n : t_n$ or $\overline{u : t}$. We write $\mathsf{dom}(\Gamma)$ for the domain of $\Gamma$, $\emptyset$ for the empty type environment, $\Gamma(u)$ for the type associated with $u$ in $\Gamma$ when $u \in \mathsf{dom}(\Gamma)$ and $\Gamma_1, \Gamma_2$ for the union of $\Gamma_1$ and $\Gamma_2$ when $\mathsf{dom}(\Gamma_1) \cap \mathsf{dom}(\Gamma_2) = \emptyset$.

If two processes share the same name $u$, they must be typed using type environments whose respective domains are not disjoint. For example, if $P$ uses $u$ according to the type $t$ and $Q$ uses $u$ according to the type $s$, they will be typed in type environments containing the associations $u : t$ and $u : s$ respectively. Overall, $u$ is used by $P \mathbin{\&} Q$ according to the type $t \cdot s$. To account for this possibility, we need to define a way of combining type environments that is more flexible than disjoint union.

**Definition 8** (type environment combination). The *combination* of $\Gamma_1$ and $\Gamma_2$ is the type environment $\Gamma_1 \cdot \Gamma_2$ such that $\mathsf{dom}(\Gamma_1 \cdot \Gamma_2) = \mathsf{dom}(\Gamma_1) \cup \mathsf{dom}(\Gamma_2)$ and defined by

$$(\Gamma_1 \cdot \Gamma_2)(u) \stackrel{\text{def}}{=} \begin{cases} \Gamma_1(u) & \text{if } u \in \mathsf{dom}(\Gamma_1) \setminus \mathsf{dom}(\Gamma_2) \\ \Gamma_2(u) & \text{if } u \in \mathsf{dom}(\Gamma_2) \setminus \mathsf{dom}(\Gamma_1) \\ \Gamma_1(u) \cdot \Gamma_2(u) & \text{if } u \in \mathsf{dom}(\Gamma_1) \cap \mathsf{dom}(\Gamma_2) \end{cases}$$

In order to enforce deadlock freedom, the type system tracks *dependencies* between objects to detect potentially dangerous circularities.

**Definition 9.** A *dependency relation* $\mathfrak{D}$ is an irreflexive, symmetric and transitive relation on names. We write $\mathsf{dom}(\mathfrak{D})$ for the domain of $\mathfrak{D}$, we write $u \sim v$ for the relation $\{(u, v), (v, u)\}$ when $u \neq v$ and $\mathfrak{D}^+$ for the transitive closure of $\mathfrak{D}$.

As we have discussed in Section 3, a dependency is established between two names $u$ and $v$ if they occur in the same message, as in $u\texttt{!m}(v)$. The rationale is that the eventual use of $v$ depends on the firing of a reaction of $u$ allowing $v$ to be received. The following notion of compatibility characterizes those cases in which the combination of two dependency relations is guaranteed not to generate mutual dependencies.

**Definition 10** (compatibility and dependency union). We say that $\mathfrak{D}_1$ and $\mathfrak{D}_2$ are *compatible* if $\mathfrak{D}_1 \cap \mathfrak{D}_2 = \emptyset$ and $(\mathfrak{D}_1 \cup \mathfrak{D}_2)^+$ is irreflexive. If $\mathfrak{D}_1$ and $\mathfrak{D}_2$ are compatible, then we write $\mathfrak{D}_1 \sqcup \mathfrak{D}_2$ for $(\mathfrak{D}_1 \cup \mathfrak{D}_2)^+$. Otherwise, $\mathfrak{D}_1 \sqcup \mathfrak{D}_2$ is undefined.

In words, two dependency relations $\mathfrak{D}_1$ and $\mathfrak{D}_2$ are (mutually) compatible if they have no dependency in common ($\mathfrak{D}_1 \cap \mathfrak{D}_2 = \emptyset$) and if their union does not generate any circular dependency by transitivity. The user of the future variable (Section 1) contains the sub-processes `user!WRITE(future)` and `future!Get(user)`. Each of them yields the dependency relation $\mathfrak{D} \stackrel{\text{def}}{=} \{(\texttt{user}, \texttt{future})\}$ and $\mathfrak{D} \cap \mathfrak{D} = \mathfrak{D} \neq \emptyset$. This suggests that the parallel composition of `user!WRITE(future)` and `future!Get(user)` is dangerous, for the respective dependency relations are incompatible. A similar thing happens between $\texttt{obj}_1$ and $\texttt{obj}_2$ in Section 3. To see the reason why it is important to also check whether $(\mathfrak{D}_1 \cup \mathfrak{D}_2)^+$ is irreflexive, consider the parallel composition

$$\big(a\,!\,\texttt{m}(b) \mathbin{\&} c\,!\,\texttt{m}(d)\big) \mathbin{\&} \big(a\,!\,\texttt{m}(c) \mathbin{\&} b\,!\,\texttt{m}(d)\big)$$

and observe that the four messages (read from left to right) respectively yield the dependency relations $\mathfrak{D}_1 \stackrel{\text{def}}{=} \{(a,b)\}$, $\mathfrak{D}_2 \stackrel{\text{def}}{=} \{(c,d)\}$, $\mathfrak{D}_3 \stackrel{\text{def}}{=} \{(a,c)\}$ and $\mathfrak{D}_4 \stackrel{\text{def}}{=} \{(b,d)\}$. Note that $(\mathfrak{D}_1 \sqcup \mathfrak{D}_2) \cap (\mathfrak{D}_3 \sqcup \mathfrak{D}_4) = \emptyset$, meaning that the two top-level sub-processes do not share any common dependency. However, if we combine all the relations together we obtain $(a,a) \in ((\mathfrak{D}_1 \sqcup \mathfrak{D}_2) \cup (\mathfrak{D}_3 \sqcup \mathfrak{D}_4))^+$. Indeed, $\mathfrak{D}_1 \sqcup \mathfrak{D}_2$ and $\mathfrak{D}_3 \sqcup \mathfrak{D}_4$ are incompatible because the transitive closure of their union is not irreflexive.

We proceed describing the typing rules, shown in Table 3. We start from the typing rules for molecules, which derive judgments of the form $\Gamma \vdash M :: t$ where $\Gamma$ collects the associations for all the arguments of the messages in $M$ and $t$ is the overall type of the molecule, intended as the product of the message types of each message in it. The rules [T-MSG-M] and [T-COMP-M] are straightforward, but for two caveats. First, the side condition in [T-MSG-M] makes sure that each message argument has a usable type. Without this condition, which is crucial for the soundness of the type system, it would be possible to have well-typed processes that violate protocols by exploiting the property that $\mathbb{0}$ is the absorbing element of the $\cdot$ connective. A detailed account of this phenomenon is given by Crafa and Padovani [16]. The second caveat is that the two rules enforce a one-to-one correspondence between names in $\Gamma$ and arguments of $M$, meaning that there cannot be two equal arguments in the same molecule. Note in particular the disjoint union $\Gamma_1, \Gamma_2$ (as opposed to the combination $\Gamma_1 \cdot \Gamma_2$) in the conclusion of [T-COMP-M]. This is a restriction compared to the original version of the type system [16] that is necessary to enforce deadlock freedom (*cf.* Section 3).

The judgments for processes have the form $\Gamma \vdash P \bullet \mathfrak{D}$, meaning that $P$ is well typed in $\Gamma$, which collects the names used by $P$, and yields the dependencies described by $\mathfrak{D}$. According to [T-DONE], the terminated process is well typed in the empty environment (it uses no names) and yields no dependencies.

Rule [T-SEND] deals with processes of the form $u\,!\,M$. The premise $\Gamma \vdash M :: t$ verifies that the molecule $M$ is well typed in $\Gamma$ and establishes its type $t$, which describes how $u$ is used in this process. The rule also establishes a dependency between the target of the message $u$ and each of the arguments occurring in $M$. Notice the disjoint union $u : t, \Gamma$ in the conclusion of the rule, indicating that $u$ cannot occur as any of the arguments in $M$, thereby excluding a self-dependency on $u$ (*cf.* Section 3).

Rule [T-PAR] deals with parallel compositions $P_1 \mathbin{\&} P_2$. Each $P_i$ is required to be well typed in a type environment $\Gamma_i$ and yields the dependencies in $\mathfrak{D}_i$. The overall composition is well typed in the composition $\Gamma_1 \cdot \Gamma_2$, meaning that in general $P_1$ and $P_2$ are allowed to share the names of objects on which they operate concurrently. Recall

**Table 3** Typing rules.

**Typing rules for processes** $\boxed{\Gamma \vdash P \bullet \mathfrak{D}}$

$$\text{[T-WEAK]}\ \frac{\Gamma \vdash P \bullet \mathfrak{D}}{\Gamma, u : \mathbb{1} \vdash P \bullet \mathfrak{D}} \qquad \text{[T-DONE]}\ \frac{}{\emptyset \vdash \texttt{done} \bullet \emptyset} \qquad \text{[T-NEW]}\ \frac{a : t \vdash C :: \mathsf{X} \qquad \Gamma, a : s \vdash P \bullet \mathfrak{D}}{\Gamma \vdash \texttt{new}\ a : t = [C]\ \texttt{in}\ P \bullet \mathfrak{D} \setminus a}\ \begin{array}{l} t \leqslant s \\ \mathsf{live}(t, \mathsf{X}) \end{array}$$

$$\text{[T-SEND]}\ \frac{\Gamma \vdash M :: t}{u : t, \Gamma \vdash u!M \bullet \bigsqcup_{v \in \mathsf{dom}(\Gamma)} u \sim v} \qquad \text{[T-PAR]}\ \frac{\Gamma_1 \vdash P_1 \bullet \mathfrak{D}_1 \qquad \Gamma_2 \vdash P_2 \bullet \mathfrak{D}_2}{\Gamma_1 \cdot \Gamma_2 \vdash P_1 \,\&\, P_2 \bullet \mathfrak{D}_1 \sqcup \mathfrak{D}_2}$$

**Typing rules for molecules** $\boxed{\Gamma \vdash M :: t}$

$$\text{[T-MSG-M]}\ \frac{}{\overline{u : t} \vdash \mathtt{m}(\overline{u}) :: \mathtt{m}(\overline{t})}\ \overline{t}\ \text{usable} \qquad \text{[T-COMP-M]}\ \frac{\Gamma_1 \vdash M_1 :: t_1 \qquad \Gamma_2 \vdash M_2 :: t_2}{\Gamma_1, \Gamma_2 \vdash M_1 \,\&\, M_2 :: t_1 \cdot t_2}$$

**Typing rules for patterns** $\boxed{\Gamma \vdash J :: \mathsf{A}}$

$$\text{[T-MSG-P]}\ \frac{}{\overline{x : t} \vdash \mathtt{m}(\overline{x}) :: \langle \mathtt{m}(\overline{t}) \rangle}\ \overline{t}\ \text{usable} \qquad \text{[T-COMP-P]}\ \frac{\Gamma_1 \vdash J_1 :: \mathsf{A}_1 \qquad \Gamma_2 \vdash J_2 :: \mathsf{A}_2}{\Gamma_1, \Gamma_2 \vdash J_1 \,\&\, J_2 :: \mathsf{A}_1 \uplus \mathsf{A}_2}$$

**Typing rules for classes** $\boxed{a : t \vdash C :: \mathsf{X}}$

$$\text{[T-REACTION]}\ \frac{\overline{x : t} \vdash J :: \mathsf{A} \qquad \overline{x : s}, a : s_0 \vdash P \bullet \mathfrak{D}}{a : t_0 \vdash J \blacktriangleright P :: \{\mathsf{A}\}}\ \begin{array}{l} \overline{t \leqslant s} \\ t_0 \leqslant t_0[\mathsf{A}] \cdot s_0 \\ t_0 \downarrow \mathsf{A} \end{array} \qquad \text{[T-CLASS]}\ \frac{a : t \vdash C_i :: \mathsf{X}_i \ ^{(1 \le i \le 2)}}{a : t \vdash C_1 \mid C_2 :: \mathsf{X}_1 \cup \mathsf{X}_2}$$

**Typing rule for solutions** $\boxed{\vdash \mathscr{D} \Vdash \mathscr{P}}$

$$\text{[T-SOLUTION]}\ \frac{a_i : t_i \vdash C_i :: \mathsf{X}_i \ ^{(i \in I)} \qquad \{a_i : s_i\}_{i \in I} \vdash \&_{j \in J} P_j \bullet \mathfrak{D}}{\vdash \{a_i : t_i = C_i\}_{i \in I} \Vdash \langle P_j \rangle_{j \in J}}\ \begin{array}{l} \forall i \in I : t_i \leqslant s_i \\ \forall i \in I : \mathsf{live}(t_i, \mathsf{X}_i) \end{array}$$

from Definition 10 that $\mathfrak{D}_1 \sqcup \mathfrak{D}_2$ is defined provided that $\mathfrak{D}_1$ and $\mathfrak{D}_2$ are compatible. Therefore, [T-PAR] is applicable only in this case.

Rule [T-NEW] deals with object definitions of the form $\mathtt{new}\ a : t = [C]\ \mathtt{in}\ P$. The premise $\Gamma, a : s \vdash P \bullet \mathfrak{D}$ checks that the continuation $P$ is well typed in a type environment enriched with an association $a : s$ for the object being defined. Notice that $t$ (the type of the object as declared in the program) and $s$ (the type according to which the object is used by $P$) are not required to be equal. It suffices that $P$ conforms with $t$, which is expressed by the side condition $t \leqslant s$. The premise $a : t \vdash C :: \mathsf{X}$ checks that the object's class is consistent with the object's type. We will describe the typing rules for classes in a moment. For the time being, it suffices to know that the set $\mathsf{X}$ collects the multisets of tags corresponding to the patterns in $C$ and is used to understand which valid configurations of $t$ are capable of triggering a reaction by means of the side condition $\mathsf{live}(t, \mathsf{X})$. The predicate $\mathsf{live}(t, \mathsf{X})$ is defined thus:

**Definition 11** (live object)**.** For all $t$ and sets of configurations $\mathsf{X}$ we have

$$\mathsf{live}(t, \mathsf{X}) \overset{\text{def}}{\iff} \forall \mathsf{A} \in [\![t]\!] : (\nexists \mathsf{B} \in \mathsf{X} : \mathsf{B} \subseteq \mathsf{A}) \Rightarrow \forall \mathtt{m}(\bar{s}) \in \mathsf{A} : \overline{s \leqslant \mathbb{1}}$$

The meaning of $\mathsf{live}(t, \mathsf{X})$ is simpler than its definition suggests. In practice, $\mathsf{live}(t, \mathsf{X})$ holds if, for any message $\mathtt{m}(\bar{s})$ of a valid configuration $\mathsf{A}$ of $t$ that may become "junk" (because $\mathsf{A}$ is unable to fire one of the reactions of the object), there are no unfulfilled obligations with respect to any of its arguments ($\overline{s \leqslant \mathbb{1}}$). Note that, when $\mathsf{A}$ is the empty configuration, there is obviously no reaction that can fire (all patterns match at least one message) and the predicate holds vacuously. Example 3 illustrates the liveness predicate for the `future` object of Section 1.

To complete the description of [T-NEW], the dependency relation $\mathfrak{D} \setminus a$ yielded by a definition for $a$ is simply the restriction of $\mathfrak{D}$ to $\mathsf{dom}(\mathfrak{D}) \setminus \{a\}$.

The last typing rule for processes is [T-WEAK], whose only purpose is to introduce in the type environment an association for a name $u$ which is not used by a process. We know that $u$ is not used because the disjoint union $u : \mathbb{1}$ with $\Gamma$ prevents $u$ from being in $\mathsf{dom}(\Gamma)$. The type of $u$ is necessarily $\mathbb{1}$, given that $u$ is not used. For the same reason, the dependencies yielded by $P$ do not change. This rule is useful in combination with [T-NEW] and [T-REACTION], which make assumptions on the presence of given associations in the type environment.

The typing rules [T-MSG-P] and [T-COMP-P] for patterns derive judgments of the form $\Gamma \vdash J :: \mathsf{A}$, meaning that $J$ matches the configuration of messages $\mathsf{A}$ and binds the arguments in $\Gamma$. The side condition in [T-MSG-P] checks that message arguments are usable, for the same reason discussed earlier for [T-MSG-M]. The two rules enforce variable linearity, ensuring that a pattern cannot bind the same argument twice.

The typing rules for classes derive judgments of the form $a : t \vdash C :: \mathsf{X}$, meaning that the class $C$ of $a$ is consistent with the type $t$ of $a$. As we have anticipated, $\mathsf{X}$ collects the configurations matched by the patterns in the reactions of $C$. The important rule is [T-REACTION], [T-CLASS] being straightforward. A reaction $J \blacktriangleright P$ of an object $a$ is well typed if so is $P$. The premise $\overline{x : t} \vdash J :: \mathsf{A}$ extracts from the pattern $J$ the environment $\overline{x : t}$ of arguments bound by $J$ and the configuration $\mathsf{A}$ of messages matched by $J$. The process $P$ is checked in a type environment that contains associations $\overline{x : s}$ for the arguments and one association $a : s_0$ for $a$. That is, $P$ is allowed to use the very same object it belongs to, pretty much as Java methods are allowed to use `this`. The side condition $\overline{t \leqslant s}$ checks that $P$ uses the arguments $\overline{x}$ conforming to their types. The side condition $t_0 \leqslant t_0[\mathsf{A}] \cdot s_0$ performs an analogous check, but concerns $a$ itself and is a little more involved. To grab the essence of the condition, recall that a reaction consumes a configuration $\mathsf{A}$ of messages among those targeted to $a$ and spawns a copy of $P$ which, in general, will produce new messages targeted to $a$ according to the type $s_0$. The reaction must maintain the overall configuration of messages targeted to $a$ valid according to the type $t_0$ of $a$, whence the side condition: after the reaction has fired and the $\mathsf{A}$ messages have been consumed, the remaining messages targeted to $a$ will be in a configuration described by $t_0[\mathsf{A}]$. This type is combined with $s_0$, which describes all the configurations of messages that $P$ may target to $a$. The resulting type must be a supertype of $t_0$. The last side condition $t_0 \downarrow \mathsf{A}$ makes sure that the environment $\overline{x : t}$ correctly reports the type of the received arguments, considering that within $t_0$ there may be different message types with the same tag and arity. The predicate is formalized thus:

**Definition 12.** $t \downarrow \langle \mathtt{m}_i(\overline{t}_i) \rangle_{i \in I} \overset{\text{def}}{\iff} \{\langle \mathtt{m}_i(\overline{s}_i) \rangle_{i \in I} \mid \langle \mathtt{m}_i(\overline{s}_i) \rangle_{i \in I \cup J} \in [\![ t ]\!]\} = \{\langle \mathtt{m}_i(\overline{t}_i) \rangle_{i \in I}\}$.

In words $t \downarrow \mathsf{A}$ holds if, given a pattern of an object of type $t$ that matches a configuration $\mathsf{A}$ of messages, the type of the arguments of the messages is uniquely determined. As an example, suppose that $t = \mathtt{A} \cdot \mathtt{m}(s_1) + \mathtt{B} \cdot \mathtt{m}(s_2)$ where $s_1$ and $s_2$ are different types. Then $t \downarrow \langle \mathtt{m}(s_1) \rangle$ does not hold, because a pattern that matches only an $\mathtt{m}$-tagged message might receive an argument of type $s_1$ or $s_2$. On the other hand, $t \downarrow \langle \mathtt{A}, \mathtt{m}(s_1) \rangle$ holds since an $\mathtt{m}$-tagged message in combination with an $\mathtt{A}$-tagged message can only have an argument of type $s_1$. Note that $t \downarrow \mathsf{A}$ implies that $t[\mathsf{A}]$ is usable. In other words, the reaction $J \blacktriangleright P$ is not dead code.

A possibly alarming aspect of [T-REACTION] is that the dependency relation $\mathfrak{D}$ yielded by $P$ does not play any role in the conclusion of the rule. One might wonder whether the firing of the reaction could, in principle, introduce new dangerous dependencies that were not present before. To see the reason why this is not possible, it is useful to look back at the reduction rule [RED] in Table 2. Before a reaction of $a$ fires, the solution must contain a process of the form $a\,!\,M$. From [T-SEND] and the typing rules for molecules, we see that this process yields dependencies between $a$ and *all* the arguments in $M$. In other words, the dependency relation of $a\,!\,M$ before the reduction is *total*. Given that the only free names of $P$ can be $a$ and the arguments received from $M$, its dependency relation $\mathfrak{D}$ is necessarily included in a total one with the same domain. Note however that $P$ may indeed introduce new dependencies involving objects that are created by $P$ itself. These new dependencies are not visible in $\mathfrak{D}$ because they have been removed by the restriction operator $\setminus$ that we have discussed with [T-NEW]. In summary, the fact that classes cannot introduce dependencies that are not observable in their clients has two fundamental consequences on the compositionality of our approach: first, there is no need to record any information on the dependencies yielded by a class; second, classes can be compiled separately and no global program analaysis is necessary in order to ensure deadlock freedom.

The only typing rule for solutions derives judgments of the form $\vdash \mathscr{D} \Vdash \mathscr{P}$. The rule does not introduce new interesting elements. Basically, it checks that each object definition in $\mathscr{D}$ is well typed and so is the parallel composition of all the processes in $\mathscr{P}$. The side conditions are exactly those we have already discussed for [T-NEW], repeated here for all the object definitions in $\mathscr{D}$.

**Theorem 1** (soundness). *If $\emptyset \vdash P \bullet \mathfrak{D}$, then $P$ is conformant and deadlock free.*

The proof of this theorem goes through a series of standard results, including type preservation. The details are given in Appendix A.

**Example 3.** Given the type `#FutureT` $\stackrel{\text{def}}{=}$ `(EMPTY·Resolve(int)+RESOLVED(int))·*Get` it is possible to obtain a typing derivation for the composition of the `future` object in Section 1 and the `user` object defined as

```
new user : #UserT [ Reply(n) ▸ System!Print(n) ]
in future!Resolve(42) & future!Get(user)
```

which resolves `future` and simultaneously attempts to retrieve its value.

In particular, for the `future` object we have that live(`#FutureT`, X) holds, where X = {⟨`EMPTY`, `Resolve(int)`⟩, ⟨`RESOLVED(int)`, `Get(Reply(int))`⟩}. Indeed, the only valid configuration of `#FutureT` that is unable to trigger a reaction of `future` is ⟨`RESOLVED(int)`⟩ and `int` is (assumed to be) irrelevant. There is only one dependency between `future` and `user` established by the message `future!Get(user)`. Therefore, according to Theorem 1, the program is both conformant and deadlock free. Note that, when the execution of the program halts, the message `future!RESOLVED(42)` is left unconsumed. According to Definition 6, the presence of this message is not a violation of deadlock freedom because `#FutureT[RESOLVED]` is irrelevant. In other words, `#FutureT` specifies that a message configuration consisting of a lone `RESOLVED` message is legal for `future`, even if this configuration is unable to trigger any reaction of `future`. ■

## 5 More Examples

In this section we show a few more well-typed programs that can be successfully analyzed with the tool Cobalt**Blue** that implements our type system. In order to deal with more interesting examples, we introduce some syntactic sugar on top of the pure Objective Join Calculus to model classes and synchronous operations.

### 5.1 Classes

A *class* is modeled as an object with a `New` operation that creates instances of the class it represents. For example, below is the class of future variables:

```
class Future [ New(r) ▸
  new this : #Future
  [ EMPTY        & Resolve(n) ▸ this!RESOLVED(n)
  | RESOLVED(n) & Get(user)  ▸ user!Reply(n) & this!RESOLVED(n) ]
```

```
5   in this!EMPTY & r!Reply(this)
6 ]
```

The code on line 5 effectively corresponds to the constructor of the class, which initializes the instance to the `EMPTY` state and sends a reference to the instance back to the object `r` that issued the request. Besides the improved readability, the advantage of using a dedicated `class` keyword instead of defining `Future` as a plain object is that we inform Cobalt**Blue** that `Future` is a *stateless* object: its only reaction has a simple join pattern involving only one message, meaning that the use of `Future` cannot contribute in creating any significant dependency between objects. This helps keeping dependency relations small, reducing the possibility that mutual dependencies arise.

### 5.2 Synchronous Operations

In addition to asynchronous message passing, Cobalt**Blue** provides syntactic sugar for synchronous operations, namely messages from which we expect to receive a result. As an example, we can model a deadlocking `user` with synchronous operations thus:

```
let future = Future.New in future!Resolve(future.Get)
```

This code creates an instance `future` of the class `Future` and attempts to resolve `future` with the value it contains. Notice the use of the '`.`' instead of '`!`' when we send a message and we expect to receive a result. This code is automatically desugared to:

```
1 new cont_1 [ Reply(future) ▸
2   new cont_2 [ CLOSURE(future) & Reply(n) ▸ future!Resolve(n) ]
3   in cont_2!CLOSURE(future) & future!Get(cont_2) ]
4 in Future!New(cont_1)
```

The output of `New` targeted to `Future` is now at the bottom (line 4) and contains the reference to a continuation object $\texttt{cont}_1$ which will receive the fresh instance of the future variable (line 1). When this happens, a `Get` message is sent to `future` along with another continuation object $\texttt{cont}_2$ to retrieve the value from the future variable (line 3). Since the continuation also needs to access the future variable, we store a reference to it into a `CLOSURE` message. When (hypothetically) the future variable returns its content, the reaction on line 2 fires and `future` is resolved. By looking at the desugared version of the code it is clear where the type checker spots the problem: both outputs on line 3 involve $\texttt{cont}_2$ and `future`. Hence, the intersection of their dependency relations is not empty – it contains the pair ($\texttt{cont}_2$, `future`) – and their parallel composition is ill typed. A well-typed variant of the program is

```
let future = Future.New in
future!Resolve(42) & System!Print(future.Get) & System!Print(future.Get)
```

where the future variable is resolved and queried by three independent processes.

In addition to synchronous messages, Cobalt**Blue** supports conditional processes, built-in arithmetic operators, anonymous object definitions and other extensions that facilitate writing more complex examples. We will make use of some of these extensions in the rest of the section, bearing in mind that they are all translated into the core language presented in Section 2 before type checking.

## 5.3 Gregory–Leibniz Approximation of $\pi$

We revisit a frequently used case study for the actor model which concerns the computation of a finite approximation of $\pi$ using the Gregory–Leibniz series:

$$\pi = 4\sum_{n=0}^{\infty}\frac{(-1)^n}{2n+1} = 4\left(1-\frac{1}{3}+\frac{1}{5}-\frac{1}{7}+\cdots\right)$$

To compute an approximation of $\pi$ using this series, we devise a network of worker processes, each responsible for a finite range of the series. We arrange the workers as a binary tree so that each leaf takes care of computing a single term of the series whereas each inner node collects the partial results from its children and sends their sum back to its parent. Below is the implementation of the `Worker` class:

```
1  class Worker [ New(depth, from, parent) ▸
2    new this : #Worker
3    [ LEAF(n, parent) ▸ parent!Reply(4. × (1 - (n % 2) × 2) / (2 × n + 1))
4    | BRANCH(parent) & Left(x) & Right(y) ▸ parent!Reply(x + y) ]
5    in if depth = 0 then
6         this!LEAF(from, parent)
7       else
8         this!BRANCH(parent) &
9         let half = from + Number.Pow(2, depth - 1) in
10        Worker!New(depth - 1, from, [ Reply(v) ▸ this!Left(v)  ]) &
11        Worker!New(depth - 1, half, [ Reply(v) ▸ this!Right(v) ])
12 ]
13 System!Print(Worker.New(10, 0))
```

The constructor is parametric in the `depth` of the tree to be built and the initial index `from` of the series to compute. A worker is either a `LEAF` or a `BRANCH`. A `LEAF` computes the term of the series corresponding to index `n`, sends the result to its parent and terminates (line 3). A `BRANCH` waits for partial results `x` and `y` from *both* its left and right children (note the use of a three-way pattern). As soon as it has collected all the necessary information, it notifies `parent` and terminates (line 4). The body of the constructor determines whether the worker is a leaf or a branch and sets its state accordingly (lines 5–11). In the latter case, two children are spawned and assigned appropriate ranges of the series to compute (lines 10–11). A critical aspect here is that children answer their parent using a `Reply` message, whereas branches expect to receive `Left` and `Right` messages. For this reason, the two children are not connected directly with their parent branch, but rather with two anonymous objects whose only purpose is to translate the `Reply` message coming from each child to a `Left` or `Right` message directed to the parent, as appropriate. The last line creates a network with 2047 workers split into 1024 leaves and 1023 branches.

The program is well typed by taking the following type definitions

```
type #Worker = #Leaf + #Branch + 𝟙
and  #Leaf   = LEAF(#Number, #Reply)
and  #Branch = BRANCH(#Reply)·Left(#Number)·Right(#Number)
and  #Reply  = Reply(#Number)
```

hence from Theorem 1 we deduce that, if the program halts (it does), no workers have been left behind. Indeed, `LEAF` and `BRANCH` have relevant arguments ($\texttt{\#Reply} \not\preccurlyeq \mathbb{1}$), meaning that in a terminated program there cannot be such messages around.

### 5.4 Sieve of Eratosthenes

We model the sieve of Eratosthenes as a chain of objects generating and processing a stream of natural numbers. At the leftmost end of the chain we have a `Generator` object that produces an infinite stream of natural numbers, starting from 2. At the rightmost end of the chain we have a `Printer` object that emits each number it receives on the terminal. For each received number $n$, `Printer` creates a new `Filter` object that removes from the stream all the subsequent numbers that are divided by $n$. This way, `Printer` only receives and prints prime numbers.

```
1  class Generator [ New(n, r) ▸
2    new this : #Generator
3    [ FROM(n) & Get(target) ▸ this!FROM(n + 1) & target!Reply(n, this) ]
4    in this!FROM(n) & r!Reply(this)
5  ]
6  class Filter [ New(k, source, r) ▸
7    new this : #Filter
8    [ READY(k, source) & Get(target) ▸ this!WAIT(k, source, target)
9    | WAIT(k, source, target) ▸
10       let n, source = source.Get in
11       if n % k = 0 then this!WAIT(k, source, target)
12       else this!READY(k, source) & target!Reply(n, this) ]
13   in this!READY(k, source) & r!Reply(this)
14 ]
15 class Printer [ New(source) ▸
16   new this : #Printer
17   [ RUN(source) ▸
18       let n, source = source.Get in
19       this!RUN(Filter.New(n, source)) & System!Print(n) ]
20   in this!RUN(source)
21 ]
22 Printer!New(Generator.New(2))
```

The `Generator` (lines 1–5) has a single state `FROM` containing the next number `n` in the stream. This is returned and incremented when the object receives a `Get` message from the next object in the chain (line 3). The `Printer` object (lines 15–21) also has a single state `RUN` containing a reference `source` to the object from which it receives prime numbers. It repeatedly sends `Get` messages to `source` (line 18) and installs a corresponding new `Filter` which becomes the new source of numbers for the subsequent iteration (line 19). Each `Filter` (lines 6–14) is parametric in the factor `k` it uses to filter out numbers from the stream and the `source` object from which it retrieves numbers and can be in one of two states. When in state `READY`, the filter waits for a `Get` request from the object `target` at its right. At this point, it moves to state `WAIT`, where it stays as long as the number `n` retrieved from `source` is a multiple of `k` (lines 10–11). As soon as the filter receives an `n` that is not a multiple of `k`, it notifies `target` with the number and moves back to state `READY` (line 12).

Given appropriate definitions for `#Generator`, `#Filter` and `#Printer` (see Appendix B), the program is well typed and therefore deadlock free. Notice that the chain of objects increases in size each time a new prime number is discovered and that deadlock freedom guarantees that the program will run forever.

## 6 Related Work

We provide an overview of other type-based approaches to the enforcement of deadlock freedom or similar properties in systems of interacting processes.

**Ordered I/O actions** Several type systems ensuring deadlock freedom (and sometimes stronger properties) for processes communicating through channels annotate channel types with *levels* describing the (abstract) moment in time the channel is going to be used for an I/O action. Dependency relations between levels capture the causal dependencies between these actions and the absence of circular dependencies implies deadlock freedom. The technique has been pioneered in the $\pi$-calculus by Kobayashi [26, 27, 28] and later adapted to more specific communication models including binary sessions [32], the conversation calculus [38], the linear $\pi$-calculus [33, 36, 35] and functional languages with linear communications [37]. Kobayashi [26] shows how the technique can be used for analyzing concurrent objects modeled in the $\pi$-calculus. In general, tracking dependencies between the single I/O actions (instead of whole channels or object names) allows for a more fine-grained and precise analysis. For example, the type systems described by Padovani [33] and Kobayashi and Laneve [29] can establish deadlock freedom for some classes of cyclic process networks, whereas our typing discipline can only deal with acyclic object networks. However, the more complex type structure of these works hampers compositionality. In particular, local changes to the network often have non-local effects of the typeability of the system to the point that it may be necessary to re-assign level annotations and re-type the smaller sub-systems. In particular, all of the aforementioned approaches except for [38, 29] use integer or natural numbers for representing levels and the typeability of a parallel composition depends on the absolute value of the levels in channel types.

**Deadlock-freedom by design** Multiparty session types [25, 3] promote a top-down methodology for the design of session-based communication systems guaranteeing deadlock freedom among the participants of the session. In this approach, the network topology and the sequence of interactions is described *a priori* in a specification called *global type* (roughly, a finite-state automaton whose transitions correspond to interactions) and checked for some mild consistency conditions. The specification is then projected into local types that are matched against each process corresponding to one of the participants of the session. If the local types and the participants do match, the participants are guaranteed to communicate safely and without deadlocks. This approach is viable when the whole system is being developed, but changes in the network topology or in the number of participants may require a substantial re-design of the global type. Also, deadlock freedom is guaranteed provided that participants do not interleave communications on different sessions. Hybrid approaches that relax this constraint and are based on a global ordering on sessions have also been considered [13, 14].

**Logically-inspired type systems for sessions** There is a close relationship between session types and linear logic propositions resulting in a Curry-Howard correspondence between proofs and session-based programs [11, 44, 31]. The deadlock-freedom property guaranteed by the type systems in these works is a direct consequence of their connection with linear logic, whereby each communication corresponds to a step of cut elimination. Our approach to deadlock freedom has been inspired by these works in the sense that the typing rules enforce a forest-like topology between concurrent objects without requiring any form of type annotation or global order on actions. The resulting compositionality of the type system is best illustrated by the typing rules for classes ([T-REACTION] and [T-CLASS]), which do not propagate any of the dependencies arising within a class outside the class itself. This means not only that objects can be type checked independently, but also that if the code of an object changes but its public interface does not, there is no need to re-check any of the other objects it interacts with. There are also major differences between the aforementioned session-based approaches and our own. First of all, session endpoints are linear resources, there are no races and interactions are triggered by complementary input/output actions. On the contrary, concurrent objects may be shared among several processes, there can be concurrent attempts to access/modify them and several messages may be necessary in order to trigger a reaction. Another technical difference with our approach is the formulation of the typing rule for parallel composition. In the type systems for sessions there are no explicit dependency relations and a parallel composition of two processes is well typed provided that the composed processes share at most one session. This condition is easy to check looking at the type environments of the two processes and suffices to prevent cycles in the resulting network topology. However, it is not as robust as compatibility with respect to the usual laws of structural congruence (commutativity and associativity of parallel composition). This slightly complicates the soundness proof of the type system and, more importantly, it implies that in order to type a program it may be necessary to rearrange parallel compositions so that the condition holds. This might not be always practical, if the bits that must be moved around are in modules whose source code is not available or cannot be changed.

A recent work by Balzer and Pfenning [5] relaxes session typing disciplines in such a way that a session endpoint may be shared among several processes. A shared endpoint must be *acquired* before being used in mutual exclusion and then *released*. The sharing is *manifest* in the sense that the type of a shareable endpoint is explicitly marked at the points where acquisition and release can/must occur. In our type system, sharing is also manifest in the sense that it is regulated by the $\cdot$ and $*$ type connectives. The former one allows object sharing among a fixed number of processes. For example, each inner node in the network of worker processes of Section 5.3 is shared between its two children. The $*$ connective allows for unrestricted sharing. For example, a future variable can be read by arbitrarily many users (Section 5.2). There is no need for native mechanisms of object acquisition and release because they can be modeled using structured synchronization patterns [16]. Balzer and Pfenning's type system is unable to ensure deadlock freedom when shared endpoints are involved, whereas our type system ensures deadlock freedom even in the presence of shared objects. Another key trait of Balzer and Pfenning's type system is that it enforces *equi-synchronization* on session endpoints: the state of an endpoint being released must be the same it had at the time of acquisition. This constraint is relaxed in our model thanks to the use of structured synchronization patterns. An acquisition request on a shared object is suspended until the object moves into a state where the request can be satisfied. As an example, all attempts to read an unresolved future variable remain pending until the variable moves into the `RESOLVED` state (Section 5.2).

**Go programs with shareable channels** Lange, Ng, Toninho, and Yoshida [30] present a type system for MiGo, a sub-language of Go focusing on communications and goroutines. The type system allows channels to be shared among several processes and guarantees deadlock freedom for well-typed programs. From a methodological standpoint, this type system differs significantly from our own since the typing rules perform little to no checks on the Go program being analyzed. Rather, the main purpose of the typing rules is to *synthesize* a type that describes at an abstract level the communications performed by the program. In fact, the type of a program is a term of a first-order process calculus such that liveness of the type (which is established by a subsequent model checking phase) implies liveness of the corresponding program. Therefore, even though the synthesis phase is compositional by definition, it is necessary to perform a global analysis on the type corresponding to the whole program in order to assess the absence of deadlocks. As the program is modified or expanded with new components, the corresponding type must be synthesized and model checked again. In contrast, our typing rules do not rely on any subsequent analysis, whence the compositional enforcement of deadlock freedom. An advantage of the global analysis is that it may result into better precision. For example, the modeling of the dining philosophers discussed by Lange, Ng, Toninho, and Yoshida [30] is ill-typed in our type system because it relies on a circular network topology. From a technical standpoint, Lange, Ng, Toninho, and Yoshida limit the sharing of channels among a finite number of processes. This property, called "fencing" in the paper, is key because it implies finiteness of the state space analyzed by the model checker. Our type system allows both finite and unrestricted sharing of objects by means of appropriate type connectives, as previously discussed.

**Other approaches** Bejleri, Mezini, and Eugster [7] define a sub-structural type system for the e-calculus, a process calculus inspired to the Join calculus for modeling event-driven, concurrent systems. Among the properties enforced by the type system is *progress*, which is akin to deadlock freedom. The main difference between the e-calculus and the Objective Join Calculus is that in the former calculus messages carry no arguments, thus none on the examples discussed in this paper can be modeled as such. It could be interesting to investigate whether the type system presented in this paper subsumes the one defined by Bejleri, Mezini, and Eugster as far as deadlock freedom is concerned. Perera, Lange, and Gay [39] sketch a technique for checking compatibility and compliance of concurrent objects. The technique is based on communicating automata and on notions of automata composition rather than types. Also in this case, systems where objects are created dynamically and messages with arguments are not considered.

## 7 Concluding Remarks

Programming large-scale concurrent systems is notoriously difficult and error-prone. Over the years, programmers have strived to make this activity less frustrating and more productive by adopting specific programming models and abstractions. In this work we propose a type-based approach for the compositional specification and implementation of such systems. The approach extends and refines the Typestate-Oriented Programming paradigm [17, 2, 42, 21] to a concurrent setting and is based on three key ingredients: a high-level model of concurrent objects with structured synchronization patterns [18, 19], a behavioral type system for specifying concurrent object protocols [16], and a mechanism that tracks dependencies between objects. Compositionality of the approach is guaranteed by the fact that classes can be type checked independently.

As shown by Crafa and Padovani [16], there are strong analogies between the Objective Join Calculus [19] and actors [24, 1]. In particular, actors are a special instance of concurrent objects where the synchronization patterns always match two messages, one that represents the state of the actor and is always self-inflicted by the actor with a so-called "become" operation, and another message that represents a request sent to the actor from a different one. The state message determines which requests the can be processed by the actor when it is in that state. Most of the examples in this paper use binary synchronization patterns and therefore adhere to this scheme. In fact, the actor model is the most popular programming framework to which our analysis technique can be applied [23]. Our type system also provides a solution to the quest for a behavioral typing discipline for actors posed by Fowler, Lindley, and Wadler [20].

The proof-of-concept implementation of the type system [34] shows the feasibility of our analysis technique. In the implementation, dependency graphs are conveniently represented using a disjoint-set data structure, allowing for efficient union operations and compatibility checks. Overall, the machinery required for deadlock freedom analysis adds negligible overhead compared to the original type system of Crafa and Padovani [16]. An important challenge for future work concerns the integration and application of the analysis technique described in this paper into a mainstream programming environment. Indeed, the type system makes use of exotic type connectives and a structural subtyping relation which apparently elude the capabilities of type systems commonly found in programming languages. One promising direction to investigate is to devise a variation of the type system that applies to the object code or bytecode of a program rather than its source code. This way, the type checker could be easily integrated in the development workflow as a separate, post-processing tool without requiring changes to the source language and its compiler.

**Acknowledgements** The author is grateful to the anonymous reviewers for their questions and thoughtful comments on an early version of this paper.

## A Supplement to Section 4

### A.1 Properties of Subtyping

We provide a semantic characterization of the derivative operator:

**Lemma 1.** *For all* $t$ *and* $\mathtt{M}$ *we have* $[\![ t[\mathtt{M}] ]\!] = \{\mathsf{A} \mid \exists \mathsf{A} : \langle \mathtt{M}' \rangle \uplus \mathsf{A} \in [\![ t ]\!] \wedge \mathtt{M} \approx \mathtt{M}'\}$.

*Proof.* The proof follows by an easy induction on $t$. □

This result is useful to show that the derivative of a type with respect to a sequence of message types is irrelevant of the order in which the single message types are considered, modulo type equivalence.

**Lemma 2.** *For all* $t$, $\mathtt{M}_1$ *and* $\mathtt{M}_2$ *we have* $t[\mathtt{M}_1][\mathtt{M}_2] \simeq t[\mathtt{M}_2][\mathtt{M}_1]$.

*Proof.* We conclude $[\![t[\mathtt{M}_1][\mathtt{M}_2]]\!] = \{\mathsf{A} \mid \exists \mathsf{A} : \langle \mathtt{M}'_1, \mathtt{M}'_2 \rangle \uplus \mathsf{A} \in [\![t]\!] \land \mathtt{M}_1 \approx \mathtt{M}'_1 \land \mathtt{M}_2 \approx \mathtt{M}'_2\} = [\![t[\mathtt{M}_2][\mathtt{M}_1]]\!]$ using Lemma 1. □

Another useful auxiliary result which will be used in the following shows that the derivatives of types related by subtyping are still related:

**Lemma 3.** *If $t \leqslant s$, then $t[\mathtt{M}] \leqslant s[\mathtt{M}]$.*

*Proof.* Let $\langle \mathtt{m}_i(\overline{s_i}) \rangle_{i \in I} \in [\![s[\mathtt{M}]]\!]$. From Lemma 1 we deduce $\langle \mathtt{m}_i(\overline{s_i}) \rangle_{i \in J} \in [\![s]\!]$ for some $J$ such that $J \setminus I$ is a singleton. From the hypothesis $t \leqslant s$ we deduce $\langle \mathtt{m}_i(\overline{t_i}) \rangle_{i \in J} \in [\![t]\!]$ and $\overline{s_i \leqslant t_i}$ for all $i \in J$. From Lemma 1 we conclude $\langle \mathtt{m}_i(\overline{t_i}) \rangle_{i \in I} \in [\![t[\mathtt{M}]]\!]$. □

### A.2 Properties of Dependency Relations

In this section we prove the robustness of compatibility with respect to associativity and name restriction. The first result in particular is key to be able to reason on parallel compositions of processes without worrying about the order in which they occur (Lemma 6).

**Lemma 4.** $(\mathfrak{D}_1 \sqcup \mathfrak{D}_2) \sqcup \mathfrak{D}_3 = \mathfrak{D}_1 \sqcup (\mathfrak{D}_2 \sqcup \mathfrak{D}_3)$.

*Proof.* The result is easily proved by thinking of $\mathfrak{D}_i$ as the corresponding undirected acyclic graph and observing that $\sqcup$ correspond to the disjoint graph union. □

**Lemma 5.** *Let $a \notin \mathsf{dom}(\mathfrak{D}_2)$. Then $\mathfrak{D}_1$ and $\mathfrak{D}_2$ are compatible if and only if $\mathfrak{D}_1 \setminus a$ and $\mathfrak{D}_2$ are compatible.*

*Proof.* ($\Rightarrow$) Observe that $\mathfrak{D}_1 \setminus a \subseteq \mathfrak{D}_1$, hence $\mathfrak{D}_1 \setminus a \cap \mathfrak{D}_2 \subseteq \mathfrak{D}_1 \cap \mathfrak{D}_2 = \emptyset$ and $(\mathfrak{D}_1 \setminus a \cup \mathfrak{D}_2)^+$ is acyclic if so is $(\mathfrak{D}_1 \cup \mathfrak{D}_2)^+$. ($\Leftarrow$) Using the hypothesis $a \notin \mathsf{dom}(\mathfrak{D}_2)$ we deduce $\mathfrak{D}_1 \cap \mathfrak{D}_2 = \mathfrak{D}_1 \setminus a \cap \mathfrak{D}_2 \subseteq \emptyset$. Now, suppose by contradiction that $(\mathfrak{D}_1 \cup \mathfrak{D}_2)^+$ has at least one cycle, and consider one of minimal length. Such cycle must go through $a$, for $(\mathfrak{D}_1 \setminus a \cup \mathfrak{D}_2)^+$ is acyclic, hence there exist $u$ and $v$ with $u \neq v$ such that $(u, a)$ and $(a, v)$ are edges of this path. Neither of these edges can be in $\mathfrak{D}_2$, for $a \notin \mathsf{dom}(\mathfrak{D}_2)$, hence they must belong to $\mathfrak{D}_1$. But this contradicts the hypothesis that $\mathfrak{D}_1$ was acyclic. □

### A.3 Subject Reduction

We will repeatedly use the next auxiliary result to rearrange parallel compositions of processes in such a way that the interesting components can be grouped together.

**Lemma 6.** *The following properties hold:*

1. $\Gamma \vdash P \mathbin{\&} Q \bullet \mathfrak{D}$ *if and only if* $\Gamma \vdash Q \mathbin{\&} P \bullet \mathfrak{D}$*;*
2. $\Gamma \vdash P \mathbin{\&} (Q \mathbin{\&} R) \bullet \mathfrak{D}$ *if and only if* $\Gamma \vdash (P \mathbin{\&} Q) \mathbin{\&} R \bullet \mathfrak{D}$.

*Proof.* Item 1 is obvious and item 2 is easily proven using Lemma 4. □

The next two results show that typing is preserved by heating and cooling rules. Hereafter we write $\mathfrak{D}_1 \asymp \mathfrak{D}_2$ if $\mathfrak{D}_1$ and $\mathfrak{D}_2$ are compatible according to Definition 10.

**Lemma 7.** *If* $\vdash \mathscr{D} \Vdash \mathscr{P}$ *and* $\mathscr{D} \Vdash \mathscr{P} \rightharpoonup \mathscr{D}' \Vdash \mathscr{P}'$, *then* $\vdash \mathscr{D}' \Vdash \mathscr{P}'$.

*Proof.* We reason by cases on the heating rule being applied.

[DONE] Then $\mathscr{D} = \{a_i : t_i = C_i\}_{i\in I} = \mathscr{D}'$ and $\mathscr{P} = \texttt{done}, \mathscr{P}'$ and $\mathscr{P}' = \{P_j\}_{j\in J}$. Let $\Gamma \stackrel{\text{def}}{=} \{a_i : t_i\}_{i\in I}$ and $P \stackrel{\text{def}}{=} \&_{j\in J} P_j$. From [T-SOLUTION] and Lemma 6 we deduce that $\Gamma \vdash \texttt{done} \mathbin{\&} P \bullet \mathfrak{D}$. From [T-PAR] we deduce that $\Gamma = \Gamma_1 \cdot \Gamma_2$ and $\mathfrak{D} = \mathfrak{D}_1 \cup \mathfrak{D}_2$ and $\Gamma_1 \vdash \texttt{done} \bullet \mathfrak{D}_1$ and $\Gamma_2 \vdash P \bullet \mathfrak{D}_2$ and $\mathfrak{D}_1 \asymp \mathfrak{D}_2$. From [T-DONE] we deduce that $\Gamma_1 = \emptyset$ and $\mathfrak{D}_1 = \emptyset$, therefore $\Gamma = \Gamma_2$ and $\mathfrak{D} = \mathfrak{D}_2$. We conclude with one application of [T-SOLUTION].

[NEW] Then $\mathscr{D} = \{a_i : t_i = C_i\}_{i\in I}$ and $\mathscr{P} = \texttt{new}\ a : t = [C]\ \texttt{in}\ P, \mathscr{P}''$ and $\mathscr{D}' = \mathscr{D}, a : t = C$ and $\mathscr{P}' = P, \mathscr{P}''$ where $\mathscr{P}'' = \{P_j\}_{j\in J}$ and $a \notin \mathsf{fn}(\mathscr{P}'')$. Let $\Gamma \stackrel{\text{def}}{=} \{a_i : t_i\}_{i\in I}$ and $Q \stackrel{\text{def}}{=} \&_{j\in J} P_j$. From [T-SOLUTION] and Lemma 6 we deduce that $\Gamma \vdash (\texttt{new}\ a : t = [C]\ \texttt{in}\ P) \mathbin{\&} Q \bullet \mathfrak{D}$ for some $\mathfrak{D}$. From [T-PAR] we deduce that $\Gamma = \Gamma_1 \cdot \Gamma_2$ and $\mathfrak{D} = \mathfrak{D}_1 \cup \mathfrak{D}_2$ and $\Gamma_1 \vdash \texttt{new}\ a : t = [C]\ \texttt{in}\ P \bullet \mathfrak{D}_1$ and $\Gamma_2 \vdash Q \bullet \mathfrak{D}_2$ and $\mathfrak{D}_1 \asymp \mathfrak{D}_2$. From [T-NEW] we deduce that $a : t \vdash C :: \mathsf{X}$ and $\Gamma_1, a : s \vdash P \bullet \mathfrak{D}_1'$ where $t \leqslant s$ and $\mathfrak{D}_1 = \mathfrak{D}_1' \setminus a$ and $\mathsf{live}(t, \mathsf{X})$. Using the hypothesis $a \notin \mathsf{fn}(\mathscr{P}'')$ we may assume, without loss of generality, that $a \notin \mathsf{dom}(\Gamma_2)$ and therefore $a \notin \mathsf{dom}(\mathfrak{D}_2)$. From Lemma 5 we deduce that $\mathfrak{D}_1' \asymp \mathfrak{D}_2$. Let $\Gamma' \stackrel{\text{def}}{=} (\Gamma_1, a : s) \cdot \Gamma_2$ and $\mathfrak{D}' \stackrel{\text{def}}{=} \mathfrak{D}_1' \cup \mathfrak{D}_2$ and observe that $\Gamma' = \Gamma, a : s$. We derive $\Gamma' \vdash P \mathbin{\&} Q \bullet \mathfrak{D}'$ with one application of [T-PAR]. We conclude with one application of [T-SOLUTION].

[PAR] Then $\mathscr{D} = \{a_i : t_i = C_i\}_{i\in I}$ and $\mathscr{P} = P \mathbin{\&} Q, \mathscr{P}''$ and $\mathscr{D}' = \mathscr{D}$ and $\mathscr{P}' = P, Q, \mathscr{P}''$ where $\mathscr{P}'' = \{P_j\}_{j\in J}$. We conclude immediately from [T-SOLUTION] and Lemma 6.

[JOIN] Then $\mathscr{D} = \{a_i : t_i = C_i\}_{i\in I}$ and $\mathscr{P} = a!(M_1 \mathbin{\&} M_2), \mathscr{P}''$ and $\mathscr{D}' = \mathscr{D}$ and $\mathscr{P}' = a!M_1, a!M_2, \mathscr{P}''$ where $\mathscr{P}'' = \{P_j\}_{j\in J}$. Let $\Gamma \stackrel{\text{def}}{=} \{a_i : t_i\}_{i\in I}$ and $Q \stackrel{\text{def}}{=} \&_{j\in J} P_j$. From [T-SOLUTION] and Lemma 6 we deduce that $\Gamma \vdash a!(M_1 \mathbin{\&} M_2) \mathbin{\&} Q \bullet \mathfrak{D}$ for some $\mathfrak{D}$. From [T-PAR] we deduce that $\Gamma = \Gamma_1 \cdot \Gamma_2$ and $\mathfrak{D} = \mathfrak{D}_1 \cup \mathfrak{D}_2$ and $\Gamma_1 \vdash a!(M_1 \mathbin{\&} M_2) \bullet \mathfrak{D}_1$ and $\Gamma_2 \vdash Q \bullet \mathfrak{D}_2$ where $\mathfrak{D}_1 \asymp \mathfrak{D}_2$. From [T-SEND] and [T-COMP-M] we deduce that $\Gamma_1 = a : t_1 \cdot t_2, \Gamma_{11}, \Gamma_{12}$ and $\Gamma_{1i} \vdash M_i :: t_i$ for all $1 \le i \le 2$ and $\mathfrak{D}_1 = \bigsqcup_{u\in\mathsf{dom}(\Gamma_1)} a \sim u$. Let $\mathfrak{D}_{1i} \stackrel{\text{def}}{=} \bigsqcup_{u\in\mathsf{dom}(\Gamma_{1i})} a \sim u$ and observe that $\mathfrak{D}_{11} \asymp \mathfrak{D}_{12}$ because $\mathsf{dom}(\mathfrak{D}_1) \cap \mathsf{dom}(\mathfrak{D}_2) \subseteq \{a\}$. Furthermore $\mathfrak{D}_1 = \mathfrak{D}_{11} \sqcup \mathfrak{D}_{12}$. We derive $a : t_i, \Gamma_{1i} \vdash a!M_i \bullet \mathfrak{D}_{1i}$ for all $1 \le i \le 2$ with two applications of [T-SEND] and $\Gamma \vdash (a!M_1 \mathbin{\&} a!M_2) \mathbin{\&} Q \bullet \mathfrak{D}$ with two applications of [T-PAR]. We conclude with one application of [T-SOLUTION]. □

**Lemma 8.** *If* $\vdash \mathscr{D} \Vdash \mathscr{P}$ *and* $\mathscr{D} \Vdash \mathscr{P} \rightharpoondown \mathscr{D}' \Vdash \mathscr{P}'$, *then* $\vdash \mathscr{D}' \Vdash \mathscr{P}'$.

*Proof.* We reason by cases on the cooling rule being applied.

[DONE] Then $\mathscr{D} = \{a_i : t_i = C_i\}_{i\in I} = \mathscr{D}'$ and $\mathscr{P}' = \texttt{done}, \mathscr{P}$ and $\mathscr{P}' = \{P_j\}_{j\in J}$. Let $\Gamma \stackrel{\text{def}}{=} \{a_i : t_i\}_{i\in I}$ and $P \stackrel{\text{def}}{=} \&_{j\in J} P_j$. From [T-SOLUTION] and Lemma 6 we deduce that $\Gamma \vdash P \bullet \mathfrak{D}$. We derive $\Gamma \vdash \texttt{done} \mathbin{\&} P \bullet \mathfrak{D}$ using one application of [T-DONE] and one application of [T-PAR]. We conclude with one application of [T-SOLUTION].

[NEW] Then $\mathscr{D} = \mathscr{D}', a : t = C$ and $\mathscr{P} = P, \mathscr{P}''$ and $\mathscr{P}' = \texttt{new}\ a : t = [C]\ \texttt{in}\ P, \mathscr{P}''$ where $\mathscr{D}' = \{a_i : t_i = C_i\}_{i\in I}$ and $\mathscr{P}'' = \{P_j\}_{j\in J}$ and $a \notin \mathsf{fn}(\mathscr{P}'')$. From [T-SOLUTION] and Lemma 6 we deduce that there exist $\Gamma$, $\mathfrak{D}$ and $\mathsf{X}$ such that $\mathsf{dom}(\Gamma) = \{a\} \cup \{a_i\}_{i\in I}$ and $t \leqslant \Gamma(a)$ and $t \leqslant \Gamma(a_i)$ for every $i \in I$ and $a : t \vdash C :: \mathsf{X}$ and $\Gamma \vdash P \mathbin{\&} Q \bullet \mathfrak{D}$ and $\mathsf{live}(t, \mathsf{X})$. From [T-PAR] we deduce that there exist $\Gamma_1$, $\Gamma_2$, $\mathfrak{D}_1$ and $\mathfrak{D}_2$ such that $\Gamma = \Gamma_1 \cdot \Gamma_2$

and $\mathfrak{D} = \mathfrak{D}_1 \cup \mathfrak{D}_2$ and $\Gamma_1 \vdash P \bullet \mathfrak{D}_1$ and $\Gamma_2 \vdash Q \bullet \mathfrak{D}_2$ and $\mathfrak{D}_1 \asymp \mathfrak{D}_2$. Without loss of generality, we may assume $a \notin \mathsf{dom}(\Gamma_2)$ using the hypothesis $a \notin \mathsf{dom}(\mathscr{P}'')$. Hence $\Gamma_1 = \Gamma_1', a : \Gamma(a)$. We derive $\Gamma_1' \vdash \texttt{new}\ a : t = [C]\ \texttt{in}\ P \bullet \mathfrak{D}_1 \setminus a$. From Lemma 5 we deduce $\mathfrak{D}_1 \setminus a \asymp \mathfrak{D}_2$. We conclude with one application of [T-PAR] and one application of [T-SOLUTION].

[PAR] Then $\mathscr{D} = \{a_i : t_i = C_i\}_{i\in I}$ and $\mathscr{P} = P, Q, \mathscr{P}''$ and $\mathscr{D}' = \mathscr{D}$ and $\mathscr{P}' = P \mathbin{\&} Q, \mathscr{P}''$ where $\mathscr{P}'' = \{P_j\}_{j\in J}$. We conclude immediately from [T-SOLUTION] and Lemma 6.

[JOIN] Then $\mathscr{D} = \{a_i : t_i = C_i\}_{i\in I}$ and $\mathscr{P} = a!M_1, a!M_2, \mathscr{P}''$ and $\mathscr{D}' = \mathscr{D}$ and $\mathscr{P}' = a!(M_1 \mathbin{\&} M_2), \mathscr{P}''$ where $\mathscr{P}'' = \{P_j\}_{j\in J}$. Let $\Gamma \stackrel{\text{def}}{=} \{a_i : t_i\}_{i\in I}$ and $Q \stackrel{\text{def}}{=} \&_{j\in J} P_j$. From [T-SOLUTION] and Lemma 6 we deduce that $\Gamma \vdash (a!M_1 \mathbin{\&} a!M_2) \mathbin{\&} Q \bullet \mathfrak{D}$ for some $\mathfrak{D}$. From [T-PAR] we deduce that $\Gamma = \Gamma_1 \cdot \Gamma_2$ and $\mathfrak{D} = \mathfrak{D}_1 \sqcup \mathfrak{D}_2$ and $\Gamma_1 \vdash a!M_1 \mathbin{\&} a!M_2 \bullet \mathfrak{D}_1$ and $\Gamma_2 \vdash Q \bullet \mathfrak{D}_2$ where $\mathfrak{D}_1 \asymp \mathfrak{D}_2$. From [T-PAR] we deduce that $\Gamma_1 = \Gamma_{11} \cdot \Gamma_{12}$ and $\mathfrak{D}_1 = \mathfrak{D}_{11} \sqcup \mathfrak{D}_{12}$ and $\Gamma_{1i} \vdash a!M_i \bullet \mathfrak{D}_{1i}$ for all $1 \leq i \leq 2$ where $\mathfrak{D}_{11} \asymp \mathfrak{D}_{12}$. From [T-SEND] we deduce that there exist $t_1$, $t_2$, $\Delta_{11}$ and $\Delta_{12}$ such that $\Gamma_{1i} = a : t_i, \Delta_{1i}$ and $\Delta_{1i} \vdash M_i :: t_i$ and $\mathfrak{D}_{1i} = \bigsqcup_{u\in\mathsf{dom}(\Delta_{1i})} a \sim u$ for all $1 \leq i \leq 2$. From $\mathfrak{D}_{11} \asymp \mathfrak{D}_{12}$ we deduce that $\mathsf{dom}(\Delta_{11}) \cap \mathsf{dom}(\Delta_{12}) = \emptyset$, for otherwise $\mathfrak{D}_{11} \cap \mathfrak{D}_{12}$ would not be empty, hence $\Gamma_1 = a : t_1 \cdot t_2, \Delta_{11}, \Delta_{12}$. We derive $\Delta_{11}, \Delta_{12} \vdash M_1 \mathbin{\&} M_2 :: t_1 \cdot t_2$ with one application of [T-COMP-M]. We derive $\Gamma_1 \vdash a!(M_1 \mathbin{\&} M_2) \bullet \mathfrak{D}_1$ with one application of [T-SEND]. We conclude with one application of [T-SOLUTION]. □

The next Lemma relates the type environment used for typing a pattern with that of a matching molecule. As expected, the types of the names in the environment of the molecule are in general subtypes of those in the environment of the pattern.

**Lemma 9.** *If $\Gamma \vdash J :: \mathsf{A}$ and $\Delta \vdash \sigma J :: t_1$ and $t \downarrow \mathsf{A}$ and there exists $t_2$ such that $t \leqslant t_1 \cdot t_2$ and $t_2 \not\leqslant \mathbb{0}$, then $\Delta \leqslant \Gamma$.*

*Proof.* From the typing rules for patterns and molecules we deduce that $\Gamma = \{\overline{x_i : t_i}\}_{i\in I}$ and $\mathsf{A} = \langle \mathtt{m}_i(\overline{t_i})\rangle_{i\in I}$ and $\Delta = \{\overline{a_i : s_i}\}_{i\in I}$ and $t_1 = \prod_{i\in I} \mathtt{m}_i(\overline{s_i})$. From the hypothesis $t_2 \not\leqslant \mathbb{0}$ we deduce that there exists $\langle \mathtt{m}_j(\overline{s_j})\rangle_{j\in J} \in [\![t_2]\!]$, where we may assume without loss of generality that $I \cap J = \emptyset$. From the definition of $[\![t_1 \cdot t_2]\!]$, we deduce that $\langle \mathtt{m}_i(\overline{s_i})\rangle_{i\in I\cup J} \in [\![t_1 \cdot t_2]\!]$. From the hypothesis $t \downarrow \mathsf{A}$ we deduce that whenever $\mathsf{A}_1 \uplus \mathsf{A}_2 \in [\![t]\!]$ and $\mathsf{A} \approx \mathsf{A}_1$ we have $\mathsf{A} = \mathsf{A}_1$. In particular, using the hypothesis $t \leqslant t_1 \cdot t_2$ we deduce that $\langle \mathtt{m}_i(\overline{t_i})\rangle_{i\in I\cup J} \in [\![t]\!]$ and $\overline{s_i \leqslant t_i}$ for every $i \in I$. Then $\Delta \leqslant \sigma\Gamma$. □

Because substitutions cannot merge syntactically different names, the substitution lemma is a minor auxiliary result.

**Lemma 10.** *If $\Gamma \vdash P \bullet \mathfrak{D}$ and $\sigma$ is a bijection with $\mathsf{dom}(\sigma) = \mathsf{dom}(\Gamma)$, then $\sigma\Gamma \vdash \sigma P \bullet \sigma\mathfrak{D}$.*

*Proof.* A simple induction on the derivation of $\Gamma \vdash P \bullet \mathfrak{D}$. □

The final Lemma towards subject reduction shows that typing is preserved by reductions.

**Lemma 11.** *If $\vdash \mathscr{D} \Vdash \mathscr{P}$ and $\mathscr{D} \Vdash \mathscr{P} \to \mathscr{D} \Vdash \mathscr{P}'$, then $\vdash \mathscr{D} \Vdash \mathscr{P}'$.*

*Proof.* We have $\mathscr{D} = a : t_0 = C, \mathscr{D}'$ and $J \blacktriangleright P \in C$ and $\mathscr{P} = a!\sigma J, \mathscr{P}''$ and $\mathscr{P}' = \sigma P, \mathscr{P}''$, where $\mathscr{D}' = \{a_i : t_i = C_i\}_{i\in I}$ and $\mathscr{P}'' = \{P_j\}_{j\in J}$. Let $Q \stackrel{\text{def}}{=} \&_{j\in J} P_j$.

From [T-SOLUTION] and Lemma 6 we deduce that there exist $s_0$, $\Gamma = \{a_i : s_i\}_{i\in I}$, $\mathsf{X}$ and $\mathfrak{D}$ such that:

A $a : t_0 \vdash C :: \mathsf{X}$

B $a : s_0, \Gamma \vdash a!\sigma J \,\&\, Q \bullet \mathfrak{D}$

C $t_0 \leqslant s_0$

Z $t_i \leqslant s_i$ for every $i \in I$

From A, [T-CLASS], [T-REACTION] we deduce that there exist $\overline{x}$, $\overline{t}$, $\overline{s}$, $\mathsf{A}$, $s_1$, $\mathfrak{D}'_1$ such that:

D $\overline{x : t} \vdash J :: \mathsf{A}$

E $\overline{x : s}, a : s_1 \vdash P \bullet \mathfrak{D}'_1$

F $\overline{t \leqslant s}$

G $t_0 \leqslant t_0[\mathsf{A}] \cdot s_1$

H $t_0 \downarrow \mathsf{A}$

From B and [T-PAR] we deduce that there exist $\Gamma_1$, $\Gamma_2$, $t_1$, $t_2$, $\mathfrak{D}_1$ and $\mathfrak{D}_2$ such that:

I $s_0 = t_1 \cdot t_2$

J $\Gamma = \Gamma_1 \cdot \Gamma_2$

▪ $\mathfrak{D} = \mathfrak{D}_1 \sqcup \mathfrak{D}_2$

L $a : t_1, \Gamma_1 \vdash a!\sigma J \bullet \mathfrak{D}_1$

M $a : t_2, \Gamma_2 \vdash Q \bullet \mathfrak{D}_2$

N $\mathfrak{D}_1 \asymp \mathfrak{D}_2$

If $a \notin \mathsf{fn}(Q)$ we can take $t_2 = \mathbb{1}$ using [T-WEAK]. From L and [T-SEND] we deduce that:

O $\Gamma_1 \vdash \sigma J :: t_1$

P $\mathfrak{D}_1 = \bigsqcup_{u\in\mathsf{dom}(\Gamma_1)} a \sim u$.

Let $\overline{c : r} \stackrel{\text{def}}{=} \Gamma_1$. Without loss of generality we may assume that $\sigma$ is the bijection $\{\overline{x \mapsto c}\}$. From D, G, H, I, O and Lemma 9 we deduce $\overline{r \leqslant t}$. From F and transitivity of $\leqslant$ we obtain $\overline{r \leqslant s}$, therefore $\Gamma_1 \leqslant \Gamma'_1$ where $\Gamma'_1 \stackrel{\text{def}}{=} \overline{c : s}$. From E and Lemma 10 we deduce

Q $\overline{c : s}, a : s_1 \vdash \sigma P \bullet \sigma\mathfrak{D}'_1$

where $\mathsf{dom}(\sigma\mathfrak{D}'_1) \subseteq \{\overline{c}, a\}$. From P we observe that $u, v \in \mathsf{dom}(\mathfrak{D}_1)$ implies $(u, v) \in \mathfrak{D}_1$, therefore $\sigma\mathfrak{D}'_1 \subseteq \mathfrak{D}_1$ and $\sigma\mathfrak{D}'_1 \asymp \mathfrak{D}_2$. From Q and M we derive

▪ $a : s_1 \cdot t_2, \Gamma'_1 \cdot \Gamma_2 \vdash \sigma P \,\&\, Q \bullet \sigma\mathfrak{D}'_1 \sqcup \mathfrak{D}_2$.

From D and O we deduce $t_1[\mathsf{A}] \leqslant \mathbb{1}$. From C, I and Lemma 3 we deduce $t_0[\mathsf{A}] \leqslant (t_1 \cdot t_2)[\mathsf{A}] = t_1 \cdot t_2[\mathsf{A}] + t_1[\mathsf{A}] \cdot t_2 \leqslant t_2$. From G and the fact that $\leqslant$ is a pre-congruence we obtain $t_0 \leqslant s_1 \cdot t_2$. We conclude with an application of [T-SOLUTION]. □

Subject reduction is a straightforward consequence of the previous results:

**Theorem 2.** *Let* $\Longrightarrow \stackrel{\text{def}}{=} (\rightharpoonup \cup \rightharpoondown \cup \rightarrow)^*$. *If* $\vdash \emptyset \Vdash P$ *and* $\emptyset \Vdash P \Longrightarrow \mathscr{D} \Vdash \mathscr{P}$, *then* $\vdash \mathscr{D} \Vdash \mathscr{P}$.

*Proof.* Immediate consequence of Lemma 7, Lemma 8 and Lemma 11. □

### A.4 Soundness

We begin by proving two auxiliary results concerning usable and relevant objects.

**Lemma 12.** *If* $\Gamma, u : t \vdash P \bullet \mathfrak{D}$*, then* $t$ *is usable.*

*Proof.* The result follows from a simple induction on the derivation of $\Gamma, u : t \vdash P \bullet \mathfrak{D}$. This proof makes key use of the side condition $\overline{\mathbb{0} \not\leqslant t}$ of rule [T-MSG-M]. □

**Lemma 13.** *If* $\Gamma, u : t \vdash P \bullet \mathfrak{D}$ *and* $t$ *is relevant, then* $u \in \mathsf{fn}(P)$.

*Proof.* The result follows from a simple induction on the derivation of $\Gamma, u : t \vdash P \bullet \mathfrak{D}$, noting that rule [T-DONE] cannot be applied (because it is only applicable on an empty type environment) and [T-WEAK] cannot be used to eliminate $u$ from the type environment because $t \not\leqslant \mathbb{1}$ from the hypothesis that $t$ is relevant. □

Next are the separate proofs of conformance and deadlock freedom for processes that are well-typed in the empty type environment.

**Lemma 14.** *If* $\emptyset \vdash P \bullet \mathfrak{D}$*, then* $P$ *is conformant.*

*Proof.* Consider a reduction $\emptyset \Vdash P \Longrightarrow \mathscr{D}, a : t = C \Vdash \langle a!\mathtt{m}_i(\overline{c_i})\rangle_{i\in I}, \mathscr{P}$ where $\mathscr{P} = \langle P_j \rangle_{j\in J}$. Let $Q \stackrel{\text{def}}{=} \&_{j\in J} P_j$. From Theorem 2, [T-SOLUTION] and Lemma 4 we deduce that there exist $s$, $\Gamma$, $\mathsf{X}$ and $\mathfrak{D}$ such that $a : t \vdash C :: \mathsf{X}$ and $a : s, \Gamma \vdash \&_{i\in I} a!\mathtt{m}_i(\overline{c_i}) \,\&\, Q \bullet \mathfrak{D}$ where $t \leqslant s$. From [T-PAR] we deduce that there exist $\Gamma_1$, $\Gamma_2$, $s_1$, $s_2$, $\mathfrak{D}_1$ and $\mathfrak{D}_2$ such that $\Gamma = \Gamma_1 \cdot \Gamma_2$ and $a : s_1, \Gamma_1 \vdash \&_{i\in I} a!\mathtt{m}_i(\overline{c_i}) \bullet \mathfrak{D}_1$ and $a : s_2, \Gamma_2 \vdash Q \bullet \mathfrak{D}_2$ and $s = s_1 \cdot s_2$. From [T-PAR], [T-SEND] and [T-MSG-M] we deduce that for every $i \in I$ there exist $\overline{t_i}$ such that $s_1 = \prod_{i\in I} \mathtt{m}_i(\overline{t_i})$. From Lemma 12 we deduce that $s_2$ is usable. From Lemma 3 we deduce that $t[\langle \mathtt{m}_i \rangle_{i\in I}] \leqslant \mathbb{1} \cdot s_2 \leqslant s_2$. We conclude that $t[\langle \mathtt{m}_i \rangle_{i\in I}]$ is usable. Now suppose that $t[\langle \mathtt{m}_i \rangle_{i\in I}]$ is also relevant and that $a$ does not occur in any of the $\overline{c_i}$ for $i \in I$. From transitivity of $\leqslant$ we deduce that $s_2$ is relevant as well. From Lemma 13 we conclude that $a \in \mathsf{fn}(Q)$. □

**Lemma 15.** *If* $\emptyset \vdash P \bullet \mathfrak{D}$*, then* $P$ *is deadlock free.*

*Proof.* We proceed by contradiction, assuming that there exists a reduction $\emptyset \Vdash P \Longrightarrow \mathscr{D} \Vdash \mathscr{P} (\rightharpoonup \cup \rightarrow)^* \not\rightarrow$ that violates the condition expressed in Definition 6. Without loss of generality, we may assume that the solution $\mathscr{D} \Vdash \mathscr{P}$ has been heated to the maximum temperature, meaning that in $\mathscr{P}$ there are no done processes, no compound molecules, no compound processes and no object definitions. In other words, we may assume that $\mathscr{P} = \langle a_i!\mathtt{m}_i(\overline{c_i})\rangle_{i\in I}$. From Theorem 2 we know that $\mathscr{D} \Vdash \mathscr{P}$ is well typed. In particular, from [T-SOLUTION] we deduce that $\mathscr{D} = \{a_i : t_i = C_i\}_{i\in I}$ and that there exist $\mathsf{X}_i$ such that $a_i : t_i \vdash C_i :: \mathsf{X}_i$ and $\mathsf{live}(t_i, \mathsf{X}_i)$ for all $i \in I$. Given $\Gamma \stackrel{\text{def}}{=} \{a_i : t_i\}_{i\in I}$, we also deduce that there exists $\mathfrak{D}$ such that $\Gamma \vdash \&_{i\in I} a_i!\mathtt{m}_i(\overline{c_i}) \bullet \mathfrak{D}$.

Now suppose that there exists $i \in I$ such that $t_i[\langle \mathtt{m}_k \rangle_{k\in I, a_k = a_i}] \not\leqslant \mathbb{1}$. We define $P_1 \stackrel{\text{def}}{=} \&_{k\in I, a_k = a_i} a_i!\mathtt{m}_i(\overline{c_i})$ and $P_2 \stackrel{\text{def}}{=} \&_{k\in I, a_k \neq a_i} a_i!\mathtt{m}_i(\overline{c_i})$. From Lemma 6 we deduce that there exist $\Gamma_1$, $\Gamma_2$, $s_1$, $s_2$, $\mathfrak{D}_1$, $\mathfrak{D}_2$ such that $\Gamma = a_i : t_i, (\Gamma_1 \cdot \Gamma_2)$ and $t_i = s_1 \cdot s_2$ and $\mathfrak{D} = \mathfrak{D}_1 \sqcup \mathfrak{D}_2$ and $a_i : s_h, \Gamma_h \vdash P_h \bullet \mathfrak{D}_h$ for all $1 \leq h \leq 2$. From [T-SEND], [T-COMP-M] and [T-MSG-M] we deduce that $s_1[\langle \mathtt{m}_k \rangle_{k\in I, a_k = a_i}] \leqslant \mathbb{1}$, therefore $t_i[\langle \mathtt{m}_k \rangle_{k\in I, a_k = a_i}] \leqslant s_1[\langle \mathtt{m}_k \rangle_{k\in I, a_k = a_i}] \cdot s_2 \leqslant s_2$ and $s_2 \not\leqslant \mathbb{1}$. From Lemma 13 we deduce that $a_i \in \mathsf{fn}(P_2)$. Given how $P_1$ and $P_2$ have been defined, it must be the case that $a_i \in \overline{c_j}$ for some $j \in I$ with $a_i \neq a_j$.

The next step is to deduce that $t_j[\langle \mathtt{m}_k \rangle_{k\in I, a_k = a_j}] \not\leqslant \mathbb{1}$, and we do that by contradiction too. If $t_j[\langle \mathtt{m}_k \rangle_{k\in I, a_k = a_j}] \leqslant \mathbb{1}$, then the multiset of messages targeted to $a_j$ is a valid configuration $\mathsf{A}$ of $t_j$. From $\mathsf{live}(t_j, \mathsf{X}_j)$ and the fact that the system cannot reduce any further (in particular, no reaction of $a_j$ is able to fire), we deduce that all the arguments of messages targeted to $a_j$ must have a type smaller than $\mathbb{1}$. This is absurd, because we know that at least one of them contains an occurrence of $a_i$ with a type that is *not* smaller than $\mathbb{1}$. In summary, starting from an object $a_i$ such that $t_i[\langle \mathtt{m}_k \rangle_{k\in I, a_k = a_i}] \not\leqslant \mathbb{1}$ we have found another object $a_j \neq a_i$ such that $t_j[\langle \mathtt{m}_k \rangle_{k\in I, a_k = a_j}] \not\leqslant \mathbb{1}$ and furthermore $a_i \in \overline{c_j}$. This argument can now be iterated to build a finite set $\{a_k\,!\,\mathtt{m}_k(\overline{c_k})\}_{k\in K} \subseteq \mathscr{P}$ such that for every $k \in K$ there exists $\omega(k) \in K \setminus \{k\}$ with $a_k \in \overline{c_{\omega(k)}}$. This set can be assumed to contain at least two elements. Without loss of generality, we also assume that it is a minimal set with the aforementioned properties.

Being a subset of $\mathscr{P}$, it must be the case that $\Delta \vdash \bigotimes_{k\in K} a_k\,!\,\mathtt{m}_k(\overline{c_k}) \bullet \mathfrak{D}'$ for some $\Delta$ and $\mathfrak{D}'$. In addition, having $K$ at least two elements, it must be possible to split this parallel composition into two parts $Q_1$ and $Q_2$ such that $Q_h \stackrel{\text{def}}{=} \bigotimes_{k\in K_h} a_k\,!\,\mathtt{m}_k(\overline{c_k})$ and $Q_h$ is well typed in isolation, for all $1 \leq h \leq 2$. More precisely, from [T-PAR] we deduce that there exist $\Delta_1$, $\Delta_2$, $\mathfrak{D}'_1$ and $\mathfrak{D}'_2$ such that $\Delta = \Delta_1 \cdot \Delta_2$ and $\mathfrak{D}' = \mathfrak{D}'_1 \sqcup \mathfrak{D}'_2$ and $\Delta_h \vdash Q_h \bullet \mathfrak{D}'_h$ for all $1 \leq h \leq 2$ where $\mathfrak{D}'_1 \asymp \mathfrak{D}'_2$. Now it must be the case that there exist $k_1$ and $k_2$ such that $k_h \in K_h$ and $\omega(k_h) \in K_{3-h}$ for all $1 \leq h \leq 2$, or else we would be able to find a smaller $K$ with the aforementioned properties, whereas we have chosen $K$ to be minimal. From [T-SEND], we deduce that $(a_{k_h}, a_{\omega(k_h)}) \in \mathfrak{D}'_1 \cap \mathfrak{D}'_2$, which contradicts $\mathfrak{D}'_1 \asymp \mathfrak{D}'_2$. In conclusion, $\mathscr{D} \Vdash \mathscr{P}$ satisfies the conditions expressed in Definition 6 hence $P$ is deadlock free. □

**Theorem 1.** *If $\emptyset \vdash P \bullet \mathfrak{D}$, then $P$ is conformant and deadlock free.*

*Proof.* Immediate consequence of Lemmas 14 and 15. □

## B Supplement to Section 5

The program discussed in Section 5.4 is well typed by taking the following type definitions:

```
type #Get       = Get(#Reply)
and  #Reply     = Reply(#Number,#Get)
and  #Generator = FROM(#Number)·#Get
and  #Filter    = READY(#Number,#Get)·#Get + WAIT(#Number,#Get,#Reply)
and  #Printer   = RUN(#Get)
```

## About the author

**Luca Padovani** is Associate Professor at the Computer Science Department of the University of Torino, where he leads the research group on Formal Methods for Software Development. Email: luca.padovani@unito.it